\documentclass{aa}  

\usepackage{graphicx}
\usepackage{txfonts}

\begin{document} 

\title{Investigating the hard X-ray emission from the hottest Abell
  cluster A2163 with Suzaku}

%   \subtitle{}

   \author{N. Ota \inst{1}, 
         K. Nagayoshi \inst{2},
         G. W. Pratt \inst{3}, 
         T. Kitayama \inst{4},
         T. Oshima \inst{5},
          \and
          T. H. Reiprich \inst{6}
% \fnmsep\thanks{}
          }

  \institute{Department of Physics, Nara Women's University, 
                 Kitauoyanishi-machi, Nara 630-8506, Japan\\
                    \email{naomi@cc.nara-wu.ac.jp}
\and
          Institute of Space and Astronautical Science (ISAS), Japan Aerospace Exploration Agency (JAXA), 
3-1-1 Yoshinodai, Sagamihara, Kanagawa 229-8510, Japan
\and
           Laboratoire AIM, IRFU/Service d'Astrophysique -- CEA/DSM -- CNRS -- Universit\'{e} Paris Diderot, B\^{a}t. 709, CEA-Saclay, 91191 Gif-sur-Yvette Cedex, France
\and
           Toho University, 2-2-1 Miyama, Funabashi, Chiba 274-8510, Japan
\and
           Nobeyama Radio Observatory, Minamimaki, Minamisaku, Nagano 384-1805, Japan
 \and
         Argelander Institute for Astronomy, Bonn University, Auf dem H\"{u}gel 71, 53121 Bonn, Germany
%             \thanks{} 
            }

  \date{Received **; accepted **}

\abstract
{We present the results from {\it Suzaku} observations of the
    hottest Abell galaxy cluster \object{A2163} at $z=0.2$.}
{To study the physics of gas heating in cluster mergers, we
     investigated hard X-ray emission from the merging cluster
     \object{A2163}, which hosts the brightest synchrotron radio
     halo.}
   {We analyzed hard X-ray emission spectra accumulated from
     two-pointed {\it Suzaku} observations. Non-thermal hard X-ray
     emission should result from the inverse Compton scattering of
     relativistic electrons by photons in the cosmic microwave
     background. To measure this emission, the dominant thermal
     emission in the hard X-ray band must be modeled in detail. To
     this end, we analyzed the combined broad-band X-ray data of
     \object{A2163} collected by {\it Suzaku} and {\it XMM-Newton},
     assuming single- and multi-temperature models for thermal
     emission and the power-law model for non-thermal
     emission. Comparing the non-thermal hard X-ray flux to radio
     synchrotron emission, we also estimated the magnetic field in the
     cluster. }
   { From the {\it Suzaku} data, we detected significant hard X-ray
     emission from \object{A2163} in the 12--60~keV band at the
     $28\sigma$ level (or at the $5.5\sigma$ level if a systematic
     error of the non-X-ray background model is considered).  The {\it
       Suzaku} HXD spectrum alone is consistent with the
     single-temperature thermal model of gas temperature
     $kT=14$~keV. From the {\it XMM-Newton} data, we constructed a
     multi-temperature model including a very hot ($kT=18$~keV)
     component in the North-East region. Incorporating the
     multi-temperature and the power-law models into a two-component
     model with a radio-band photon index, i.e., $\Gamma=2.18$, the
     12--60~keV energy flux of non-thermal emission is constrained
     within $5.3\pm0.9\,(\pm3.8)\times10^{-12}~{\rm
       erg\,s^{-1}cm^{-2}}$ (the first and second errors refer to the
     $1\sigma$ statistical and systematic uncertainties,
     respectively). The 90\% upper limit of detected inverse Compton emission is
     marginal ($F_{\rm NT} < 1.2\times10^{-11}~{\rm
       erg\,s^{-1}cm^{-2}}$ in the 12--60~keV band). The estimated
     magnetic field in \object{A2163} is $B > 0.098~{\rm \mu
       G}$. While the present results represent a three-fold increase
     in the accuracy of the broad band (0.3--60~keV) spectral model of
     A2163, more sensitive hard X-ray observations are needed to
     decisively test for the presence of hard X-ray emission due to inverse Compton
     emission.}
   {}

   \keywords{galaxies: clusters: individual: Abell~2163 --
     galaxies: intracluster medium -- X-rays: galaxies: clusters --
     cosmology: observations
               }

  \authorrunning{N. Ota et al.}
  \titlerunning{Hard X-ray emission from A2163}

   \maketitle
%
%________________________________________________________________

\section{Introduction}\label{sec:intro}
The most energetic events in the Universe since the Big Bang are
cluster mergers, with kinetic energy on the order $10^{65}$~ergs.
Cluster collisions release a huge amount of energy, a fraction of
which is expected to heat the gas and generate non-thermal particles
through shock waves. Hence, shock-heated gas provides important clues
for understanding high-energy phenomena and the evolution of galaxy
clusters. Synchrotron emission at radio wavelengths extending over
scales of megaparsec have been observed in many clusters
\citep{2012A&ARv..20...54F}. The existence of extended radio halos
suggest the acceleration of relativistic electrons in the intracluster
space. In the hard X-ray band, non-thermal emission is predicted to be
generated by the inverse Compton (IC) scattering of these electrons by
photons in the cosmic microwave background (CMB).

IC hard X-ray emissions from nearby clusters have been detected by
{\it RXTE} and {\it Beppo-SAX} satellites \citep[for review,
see][]{2008SSRv..tmp...16R}. \object{The Coma cluster} is the
archetypal case \citep{2002ApJ...579..587R, 2004ApJ...602L..73F}. On
the other hand, no significant non-thermal X-ray emissions have been
detected by the {\it Suzaku} and {\it Swift} satellites
\citep{2009ApJ...696.1700W, 2011ApJ...727..119W}. This discrepancy
might be reconciled by considering different sizes of viewing fields
\citep{2011ApJ...732...85F}. {\it Suzaku} has searched for non-thermal
X-ray emissions in nine bright clusters: \object{the Centaurus
  cluster} \citep{2007xnnd.confP..28K}, \object{the Ophiuchus cluster}
\citep{2008PASJ...60.1133F}, \object{RX~J1347.5--1145}
\citep{2008A&A...491..363O}, \object{A3376}
\citep{2009PASJ...61S.377K}, \object{A2319}
\citep{2009PASJ...61.1293S}, \object{A3667}
\citep{2009PASJ...61..339N}, \object{the Coma cluster}
\citep{2009ApJ...696.1700W}, \object{the Perseus cluster}
\citep{2010PASJ...62....9N}, and \object{A2199}
\citep{2010PASJ...62..115K}.  The hard X-ray spectra can be explained
by thermal emission; adding a non-thermal power-law component does not
improve data fitting \citep[][and reference
therein]{2012RAA....12..973O}.
\citet{2009ApJ...690..367A,2010ApJ...725.1688A} suggested on the basis
of the {\it Swift} observations of 20 clusters that the hard X-ray
emission from clusters (except \object{the Bullet cluster}) most
likely has a thermal origin. Stacked spectra, constructed and analyzed
from {\it Fermi} data, yielded no significant $\gamma$-ray signal from
clusters \citep{2013A&A...560A..64H}.  \citet{2013arXiv1308.5654T}
searched for cosmic-ray induced $\gamma$-ray emission through a
combined analysis of 50 clusters to exclude hadronic injection
efficiency in simple hadronic models, and also derived limits on the
$\gamma$-ray flux on individual clusters.

It is worth noting that, for merging clusters, the radio synchrotron
power $P_{1.4}$ is correlated with the X-ray luminosity of the thermal
emission $L_X$. On the other hand, relaxed clusters with no radio halo
lie in a completely separate regime in the $P_{1.4}-L_X$ plane
\citep{2009A&A...507..661B,2013arXiv1306.4379C}.  This suggested that
the generation of high-energy particles is connected to cluster
evolution. Massive clusters that emit luminous X-rays contain
high-energy particles and an intracluster magnetic field. Thus, we
question whether {\it radio-loud} clusters similarly produce
significant non-thermal X-ray emission? To answer this question, we
focus on the hottest Abell cluster \object{A2163}, located at the
brightest end of the $P_{1.4}-L_X$ relationship.

The mean temperature of \object{A2163} ($z=0.203$) is $14$~keV
\citep{1992ApJ...390..345A}. This cluster hosts a huge, powerful
synchrotron halo of radio power $P_{1.4}=155$~mJy
\citep{2004A&A...423..111F} and also possesses a complex temperature
structure \citep{2001ApJ...563...95M, 2011A&A...527A..21B}. The
presence of high-temperature gas in the cluster has been confirmed by
the {\it XMM-Newton} and {\it Chandra} observations, indicating that
the cluster has undergone recent merging. From weak lensing
observations, \citet{2011ApJ...741..116O,2012A&A...540A..61S} showed
that the mass distribution in \object{A2163} is bimodal, which
supports the merging hypothesis.  Hard X-ray observations of
\object{A2163} were carried out by {\it Beppo-SAX} and {\it
  RXTE}. {\it Beppo-SAX} yielded the 90\% upper limit of non-thermal
IC emission as $F_{\rm NT}(20-80~{\rm keV})<5.6\times10^{-12}~{\rm
  erg\,s^{-1}cm^{-2}}$ \citep[][]{2001A&A...373..106F}.  Detection was
claimed from the {\it RXTE} data with long exposure time but was
associated with large uncertainty \citep[$F_{\rm NT}(20-80~{\rm
  keV})\sim 1.1^{+1.7}_{-0.9}\times10^{-11}~{\rm
  erg\,s^{-1}cm^{-2}}$;][]{2006ApJ...649..673R}.

To constrain the non-thermal hard X-ray emission from the
\object{A2163} cluster, we analyze hard X-ray spectra obtained by the
Hard X-ray Detector \citep[HXD;][]{2007PASJ...59S..35T} onboard the
{\it Suzaku} satellite \citep{2007PASJ...59S...1M}.  From a joint
analysis of the {\it Suzaku} and {\it XMM-Newton} data, we aim to
understand the origin of hard X-ray emission and properties of
shock-heated gas.

Throughout this paper, we adopt a cosmological model with standard
parameters: matter density $\Omega_{M}=0.27$, cosmological constant
$\Omega_{\Lambda}=0.73$, and the Hubble constant $H_0=70~{\rm
  km\,s^{-1}\,Mpc^{-1}}$.  At the cluster redshift ($z=0.203$),
$1\arcmin$ corresponds to 201~kpc.  Unless otherwise specified, quoted
errors indicate the 90\% confidence intervals.

\section{Observation and data reduction}
\subsection{{\it Suzaku}/HXD}
\begin{figure}[htb]
   \centering
\rotatebox{0}{\scalebox{0.40}{\includegraphics{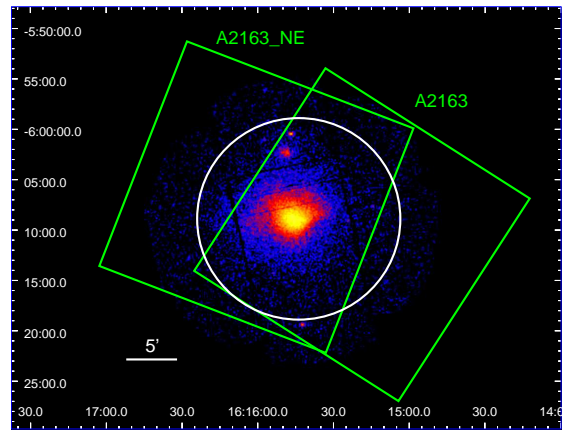}}}
\caption{{\it XMM-Newton} EMOS image of \object{A2163} in the
  0.2--10~keV band. The boxes indicate the HXD-PIN field of views
  during two pointed observations. The white circle delineates the
  {\it XMM-Newton} spectral region of cluster emission
  ($r<10\arcmin$). }
   \label{fig1}%
 \end{figure}

\object{A2163} has been observed in two pointings (Fig.~\ref{fig1}):
the central region (A2163, PI: T. Reiprich) and the North-East region
(A2163\_NE, PI: N. Ota). The observational details are listed in
Table~\ref{tab1}. The HXD-PIN spectral data have a narrow
field-of-view ($30\arcmin \times 30\arcmin$ (FWHM)) and low background
level \citep[][]{2007PASJ...59S..35T} enabling the study of hard X-ray
emission from \object{A2163} up to several tens of keV. This energy
range is ideal for this study.

%%%%%%%%%%%%%%%%%%%%%%%%%%%%%%%%%%%%%%%%%%%%
\begin{table*}
\caption{Log of {\it Suzaku} observations of \object{A2163}. }\label{tab1}
\centering
\begin{tabular}{llllll} \hline\hline
Target & Obs ID & Date  &\multicolumn{2}{c}{Coordinates$^{\mathrm{a}}$} & Exposure$^{\mathrm{b}}$ \\ 
 & & & RA  & Dec &   [s] \\ \hline
 A2163     & 803071010 & 2008 Aug 18--22 & 16:15:15.7  & $-06$:06:25.9 & 113380 \\
A2163\_NE & 803022010 & 2009 Feb 08--10 &  16:16:06.2 &	 $-06$:03:32.8 &  40846 \\\hline  
\end{tabular}
\begin{list}{}{}
\item[$^{\mathrm{a}}$] Pointing coordinates in J2000.
\item[$^{\mathrm{b}}$] HXD net exposure time after data filtering.
\end{list}
\end{table*}
%%%%%%%%%%%%%%%%%%%%%%%%%%%%%%%%%%%%%%%%%%%%
%%%%%%%%%%%%%%%%%%%%%%%%%%%%%%%%%%%%%%%%%%%%
%%%%%%%%%%%%%%%%%%%%%%%%%%%%%%%%%%%%%%%%%%%%

\begin{figure*}[htb]
   \centering
\rotatebox{0}{\scalebox{0.33}{\includegraphics{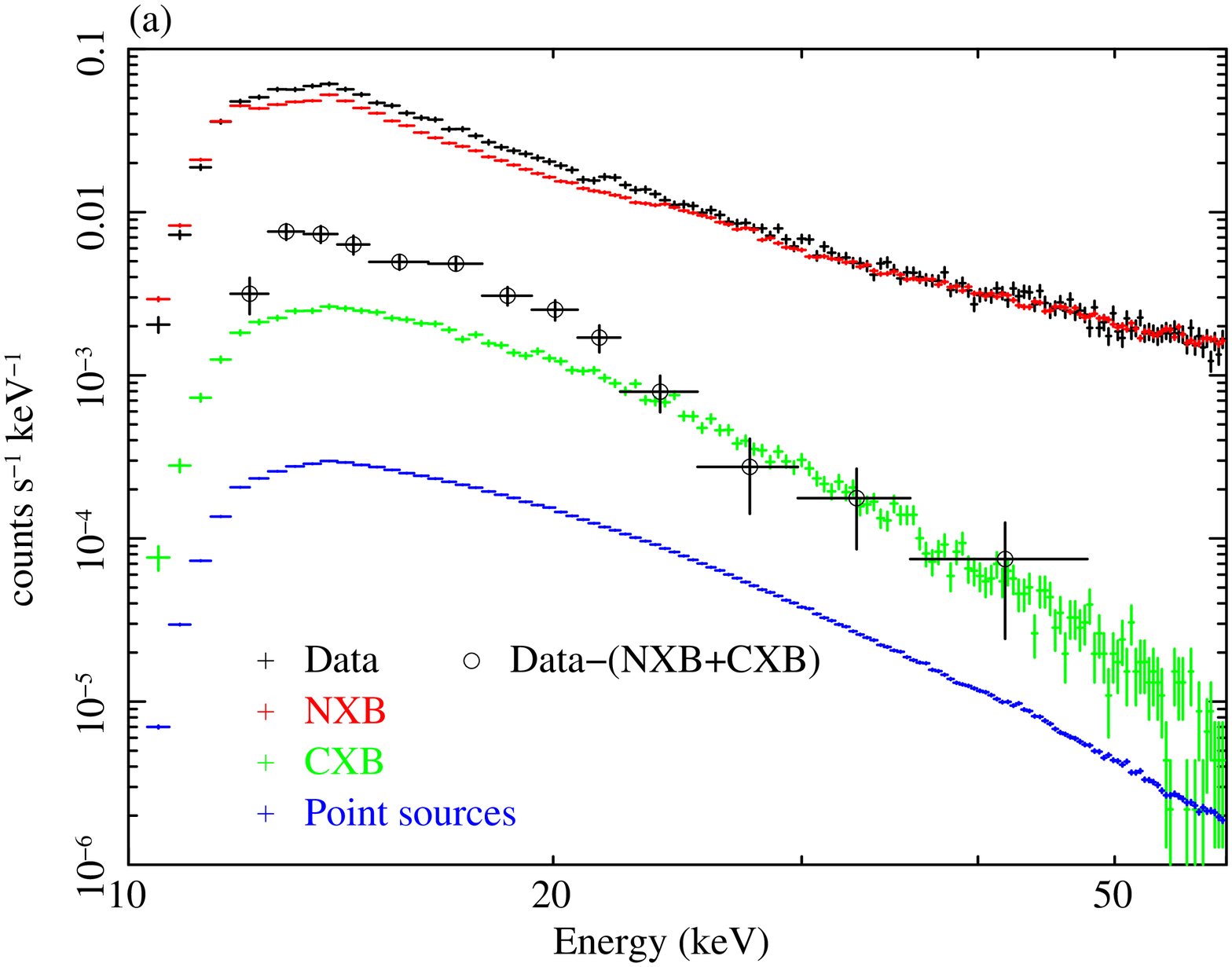}}}
\rotatebox{0}{\scalebox{0.33}{\includegraphics{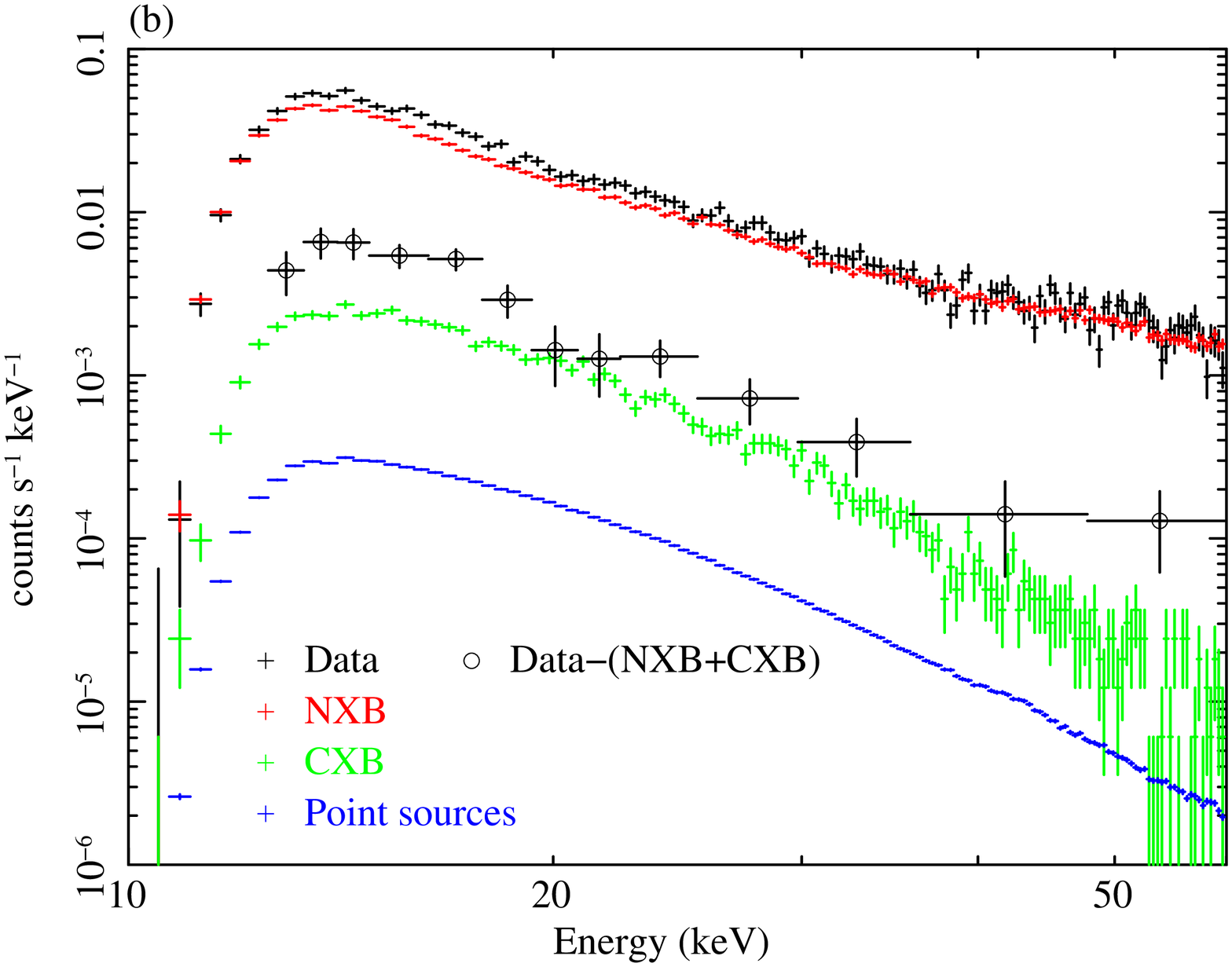}}}
\caption{HXD spectra of \object{A2163} in the 10--60~keV band for the
  central region (a) and the north-east region (b).  In each panel,
  observed HXD data denoted as ``Data'', the NXB model, the CXB model,
  and the background-subtracted data, ``Data -- (NXB+CXB)'', are shown
  using black crosses, red crosses, green crosses, and black circles,
  respectively. The point-source contribution estimated from the {\it
    XMM-Newton} observations of the same field is shown using blue
  crosses. }
   \label{fig2}%
 \end{figure*}

 HXD data reduction was performed using {\tt HEASOFT} version 6.13 and
 CALDB version 2011-09-13 for HXD.  The data were reprocessed and
 screened in a standard manner using the {\it Suzaku} reprocessing
 tool {\tt aepipeline}. The screening criteria are as follows: Earth
 elevation angle $>5^{\circ}$, geomagnetic cut-off rigidity $>6$~GV,
 satellite outside the South Atlantic anomaly.  The source spectra
 were extracted by {\tt hxdpinxbpi}. Fig.~\ref{fig2} shows the HXD
 spectrum for each pointing.

 %%%%%%%%%%%%%%%%%%%%%%%%%%%%%%%%%%%%%%%%%%%%
%%%%%%%%%%%%%%%%%%%%%%%%%%%%%%%%%%%%%%%%%%%%

 The HXD-PIN detector background was subtracted using the non-X-ray
 background (NXB) files provided by the HXD instrument team. The
 cosmic X-ray background (CXB) was calculated by a power-law model
 with an exponential cut-off at 40~keV, previously determined from
 {\it HEAO-1} A2 data \citep{1987IAUS..124..611B}. In spectral
 fitting, we used the PIN response function, which is appropriate for
 the observational epoch, but corrected for the off-axis response with
 {\tt hxdarfgen}, approximating the surface brightness profile by the
 $\beta$-model and extending it to $r=10\arcmin$.

 The background in the HXD band is dominated by NXB.  Following the
 same method described in \citet{2008A&A...491..363O}, the accuracy of
 NXB model is estimated to be 2\% based on a comparison between the
 NXB model and data collected during periods of Earth occultation.
 This estimate agrees with the typical uncertainty reported in
 \citet{2009PASJ...61S..17F}.  Thus, we assume the $1\sigma$
 systematic error = 2\% and propagate this error throughout our
 spectral analysis of HXD-PIN data.

 We also estimated a hard X-ray flux of point sources inside the HXD
 field-of-view from the {\it XMM-Newton} data of the same field (see
 the next subsection). Assuming the average photon index of 27
 detected sources $\Gamma=1.5$, we simulated the HXD spectrum expected
 for each pointing. Here the HXD angular response function was
 considered by using {\tt hxdarfgen}. In Fig.~\ref{fig2}, the spectrum
 of the sum of the detected sources is shown. The contribution from
 the point sources to the source spectrum is estimated as 6\%, which
 is negligible in comparison with the uncertainty of the
 above-mentioned NXB model.

\subsection{{\it XMM-Newton}}
\object{A2163} was observed in a mosaic of five pointings in
2000. This study analyzes only the central pointing, OBSID 0112230601,
obtained in revolution 0132. Observation data files were retrieved
from the {\it XMM-Newton}
archive\footnote{\url{http://xmm.esac.esa.int/xsa/}} and reprocessed
with the {\it XMM-Newton} Science Analysis System (SAS) v10.0 using
the standard calibration. The initial data reduction, including solar
flare screening, event selection, and vignetting correction, followed
the procedure in \citet{pra07}. The exposure time after data screening
was 10 and 6 ks for EMOS and EPN, respectively. The background data
sets were the accumulated blank-sky data of \citet{rea03}; these were
subjected to the same screening and vignetting correction as the
source files and normalized by the count rates in the 10--12 keV band
for EMOS and in the 12--14~keV band for EPN. Background subtraction
was performed in two steps, as described in \citet{arn02}, using a
source-free annulus at the edge of the field of view ($r >
11\farcm5$).

\section{HXD analysis}\label{sec:hxd_analysis}
\begin{figure*}[htb]
   \centering
\rotatebox{0}{\scalebox{0.33}{\includegraphics{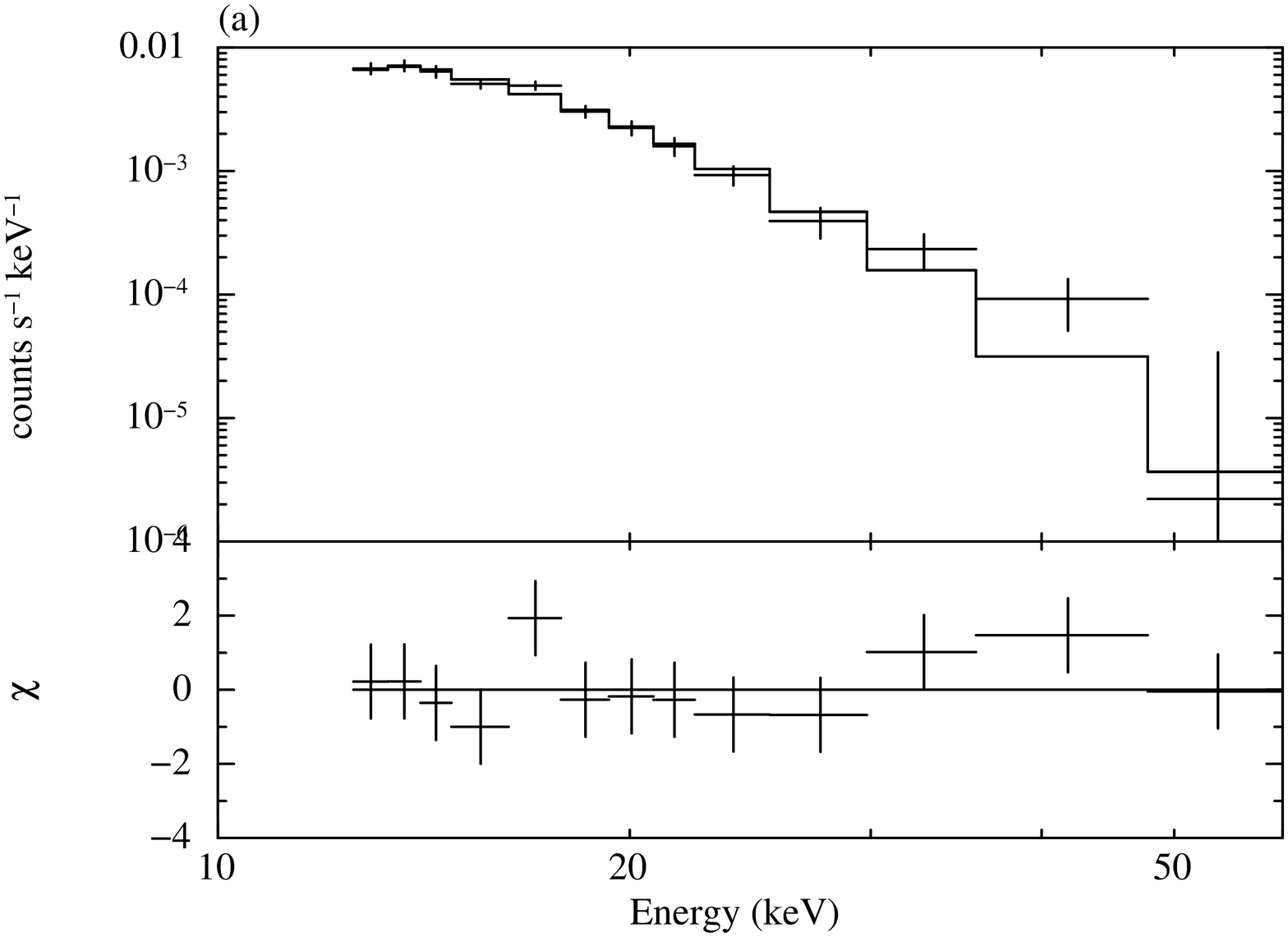}}}
\rotatebox{0}{\scalebox{0.33}{\includegraphics{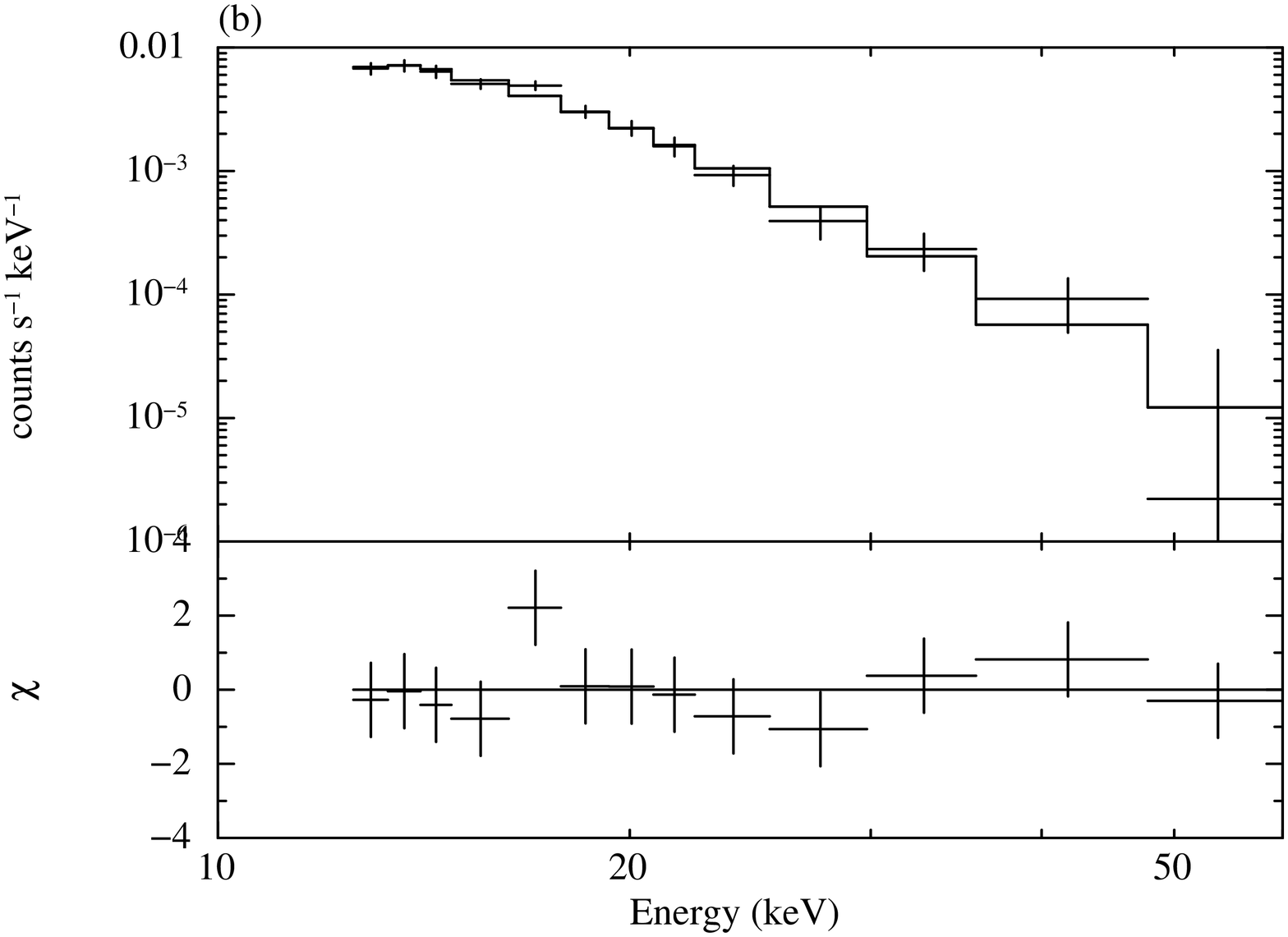}}}
\caption{HXD spectrum of \object{A2163} in the 12--60~keV band with
  CXB and NXB subtracted. The HXD spectrum (the crosses in the upper
  plots) is fitted by the APEC model (a) and the power-law model
  (b). Below the spectra, the residuals are expressed in number of
  standard deviations.}
   \label{fig3}%
 \end{figure*}

Since the spectra extracted from the two data sets are statistically
consistent across the HXD-PIN band, both spectra were added, yielding
a total exposure time of 154~ks. Figure~\ref{fig3} shows the HXD
spectrum of \object{A2163} with the NXB and CXB components
removed. The 12--60~keV flux is measured as $1.70\pm 0.06\, (\pm 0.30)
\times 10^{-11}~{\rm erg\,s^{-1}\,cm^{-2}}$, where the first and
second errors are the $1\sigma$ statistical and $1\sigma$ systematic
errors, respectively. The hard X-ray emission is detected at the
$28\sigma$ confidence level. Considering the systematic error of the
NXB, the significance of the detection is $5.5\sigma$. Hence the
present data provide the highest-quality X-ray spectra of
\object{A2163}.

Next, the HXD spectrum in the 12--60~keV band was fitted to two
single-component models; (1) the Astrophysical Plasma Emission Code
(APEC) thin-thermal plasma model \citep{2001ApJ...556L..91S} and (2)
the power-law model. The Galactic absorption was fixed at $N_{\rm H}=
1.65\times10^{21}~{\rm cm^{-2}}$. The results are shown in
Fig.~\ref{fig3} and Table~\ref{tab2}. In Model (1), the metallicity
and redshift were fixed at $Z=0.3$~solar and $z=0.203$, respectively.

To examine the impact of background uncertainty, the NXB intensity was
intentionally altered by $\pm 2$\%. We find that both models are
statistically acceptable at the 90\% level. The measured temperature
$14^{+4}_{-3} (^{+2}_{-2})$~keV and normalization factor
$3.6^{+1.1}_{-0.8}(^{+0.2}_{-0.1}) \times 10^{-2}$ obtained from Model
(1) favorably agreed with those obtained from the {\it XMM-Newton}
data, $kT=13.5^{+1.1}_{-0.8}$~keV and
Norm$=3.60^{+0.09}_{-0.08}\times10^{-2}$.  Thus, the relative
normalization factor between {\it Suzaku} HXD and {\it XMM-Newton}
EMOS1 is $1.00\pm0.28$. This result is statistically consistent with
the cross-calibration between {\it Suzaku} XIS and {\it XMM-Newton}
\citep{2011A&A...525A..25T}, considering the relative normalization
factor between {\it Suzaku} XIS and HXD reported by the instrument
team\footnote{http://www.astro.isas.jaxa.jp/suzaku/doc/suzakumemo/suzakumemo-2008-06.pdf}.
On the other hand, the power-law model requires a large photon index
of $\Gamma\sim 3$ to fit the data, indicating that the cluster
emission spectrum is {\it soft} and predominantly thermal.

%%%%%%%%%%%%%%%%%%%%%%%%%%%%%%%%%%%%%%%%%%%%
%%%%%%%%%%%%%%%%%%%%%%%%%%%%%%%%%%%%%%%%%%%%
%%%%%%%%%%%%%%%%%%%%%%%%%%%%%%%%%%%%%%%%%%%%
\begin{table}
\caption{Single-component fitting to the HXD data }\label{tab2}
\centering
\begin{tabular}{ll} \hline\hline
Parameter & Value$^{\mathrm{a}}$ \\ \hline
\multicolumn{2}{c}{APEC model} \\ \hline
$kT$ [keV] & $14^{+4}_{-3} (^{+2}_{-2})$\\
Abundance [solar] & 0.3 (Fix)\\
Redshift & 0.203 (Fix)\\
Norm$^{\mathrm{a}}$ & $3.6^{+1.1}_{-0.8}(^{+0.2}_{-0.1}) \times 10^{-2}$ \\
$\chi^2$/dof & 8.6/11\\ \hline
\multicolumn{2}{c}{Power-law model} \\ \hline
$\Gamma$ & $2.9^{+0.3}_{-0.3} (^{+0.2}_{-0.2})$ \\
Norm$^{\mathrm{b}}$ & $0.13^{+0.17}_{-0.07} (^{+0.07}_{-0.04})$\\
$\chi^2$/dof & 8.3/11\\ \hline
\end{tabular}
\begin{list}{}{}
\item[$^{\mathrm{a}}$] The first and second (in parenthesis) errors
  are the statistical and systematic errors, respectively.
\item[$^{\mathrm{b}}$] Normalization of the APEC model, $Norm = \int
  n_e n_H dV/(4\pi (1+z)^2 D_A^2)~[10^{-14}{\rm cm^{-5}}]$. $D_A$ is
  the angular diameter distance to the source. 
\item[$^{\mathrm{c}}$] Normalization of the Power-law model, in units of ${\rm
    photons\,keV^{-1}\,cm^{-2}\,s^{-1}}$ at 1~keV.
\end{list}
\end{table}
%%%%%%%%%%%%%%%%%%%%%%%%%%%%%%%%%%%%%%%%%%%%

\section{XMM+HXD joint analysis}\label{sec:joint}
To more thoroughly investigate the origin of hard X-ray emission, we
performed a joint XMM+HXD analysis.  Since most of the cluster
emission appears to be thermal, the thermal emission must be modeled
in detail to constrain the contribution from non-thermal hard X-ray
emissions. To this end, we replicate thermal emission using a
single-component APEC model and multi-temperature APEC models and
constrain the quantity of non-thermal hard X-ray emission.

\subsection{Single-temperature model}
\begin{figure*}[htb]
   \centering
\rotatebox{0}{\scalebox{0.33}{\includegraphics{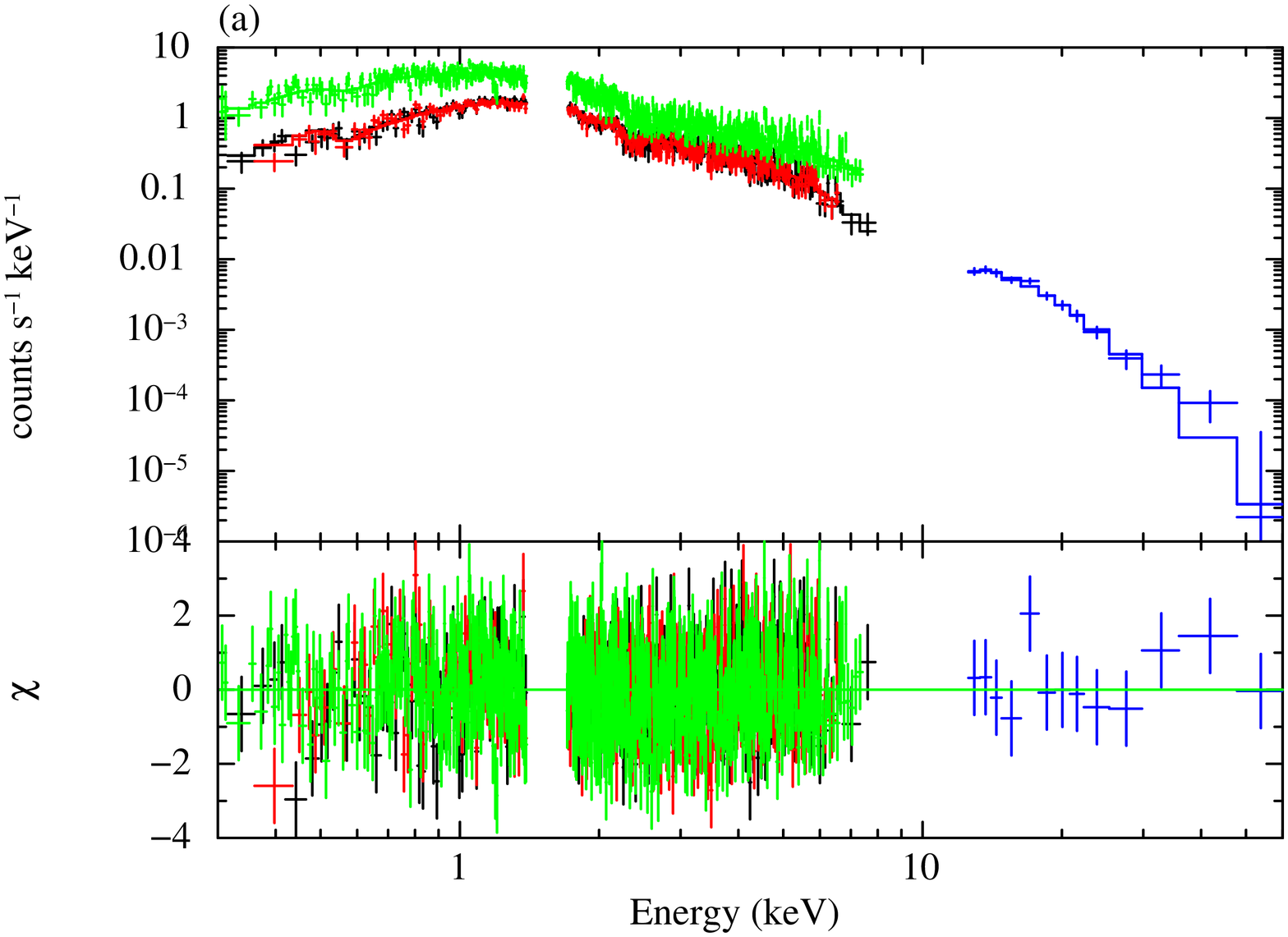}}}
\rotatebox{0}{\scalebox{0.33}{\includegraphics{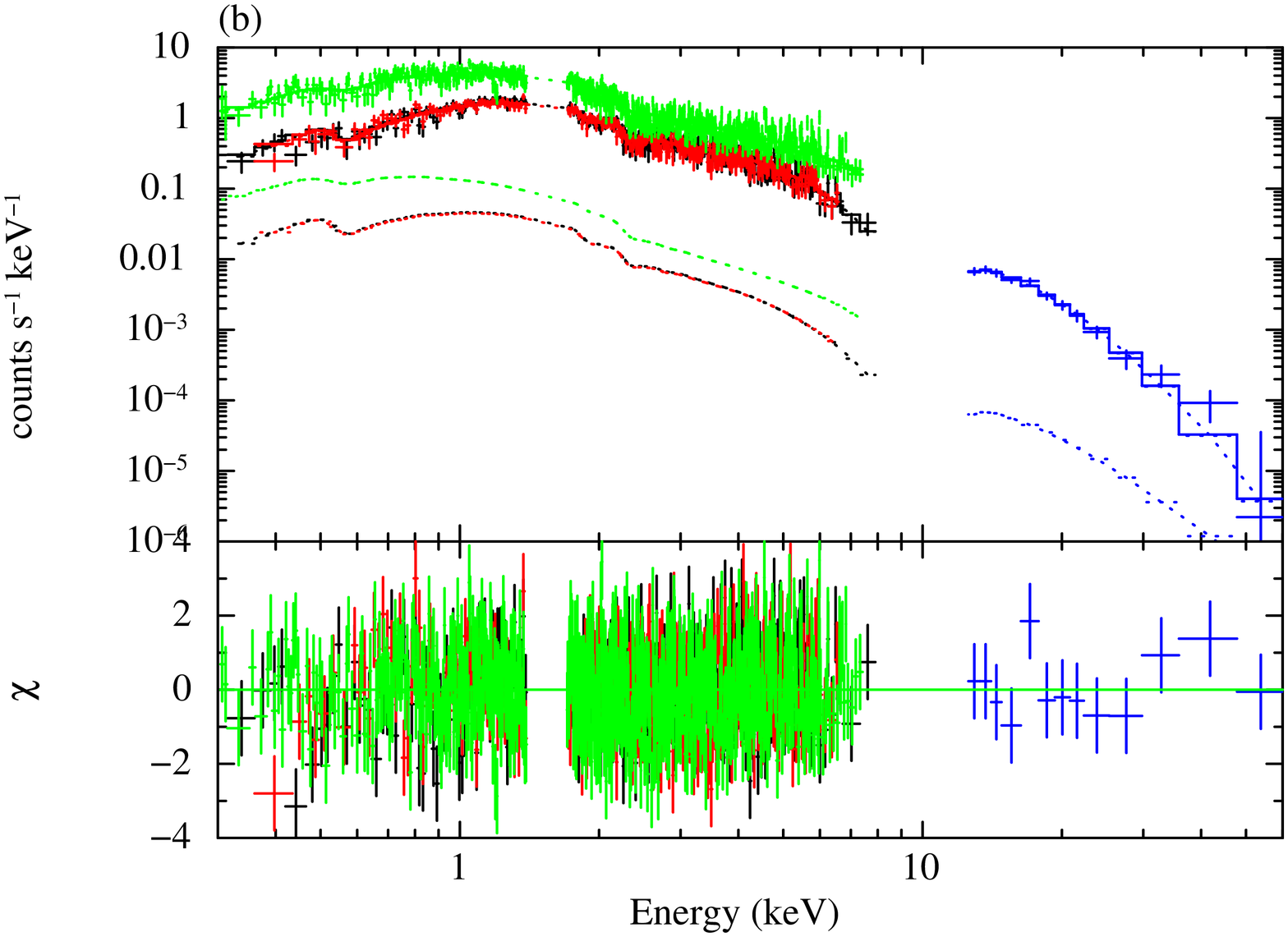}}}
\caption{XMM + HXD spectra of \object{A2163} in the 0.3--60~keV band
  fitted by (a) the APEC model and (b) the APEC+power-law model. The
  EMOS1/EMOS2/EPN data are shown by the black/red/green crosses; the
  blue crosses denote HXD data. Solid lines indicate the best-fit
  models for each instrument. In panel (b), the power-law component is
  also indicated by the dotted line. }
   \label{fig4}%
 \end{figure*}

Figure~\ref{fig4} and Table~\ref{tab3} show the results of fitting the
XMM+HXD broad-band spectra of \object{A2163} to the single-temperature
APEC model. The 0.3--60~keV spectra were well fitted by a
$kT\sim14$~keV thermal model, indicating that the observed hard X-ray
emission is likely dominated by hot thermal emission. The bolometric
luminosity was estimated as $9.0\times10^{45}~{\rm erg\,s^{-1}}$.

Incorporating the power-law component into the spectral model, we can
derive the upper limit of the non-thermal component. The power-law
index assumes a radio-band photon index; i.e., $\Gamma=2.18$
\citep{2004A&A...423..111F}. The combined model did not significantly
improve the fit relative to the case of APEC model alone; the
resulting $\chi^2/{\rm dof}=1238/1179$. Thus, we infer the absence of
significant non-thermal hard X-ray emission. Summing the statistical
and systematic errors in quadrature, we estimate the 90\% upper limit
on the 12--60~keV energy flux as $F_{\rm NT}< 1.2\times 10^{-12}~{\rm
  erg\,s^{-1}\,cm^{-2}}$.

%%%%%%%%%%%%%%%%%%%%%%%%%%%%%%%%%%%%%%%%%%%%
%%%%%%%%%%%%%%%%%%%%%%%%%%%%%%%%%%%%%%%%%%%%
%%%%%%%%%%%%%%%%%%%%%%%%%%%%%%%%%%%%%%%%%%%%
\begin{table*}
\caption{Single-temperature analysis of the XMM+HXD data}\label{tab3}
\centering
\begin{tabular}{lll} \hline\hline
                  & APEC & APEC+Power \\ \hline
$kT$ [keV] & $14.1^{+0.6}_{-0.6}\, (^{+1.2}_{-1.1})$ & $14.4^{+0.7}_{-0.7}\, (^{+1.7}_{-1.4})$ \\
Abundance [solar] & $0.29^{+0.10}_{-0.10}\, (^{+0.01}_{-0.01})$ & $0.31^{+0.11}_{-0.11}\, (^{+0.01}_{-0.01})$\\
Norm$^{\mathrm{a}}$ & $3.62^{+0.08}_{-0.08}\,(^{+0.03}_{-0.04})\times 10^{-2}$ & $3.53^{+0.12}_{-0.12}\,(^{+0.05}_{-0.06})\times 10^{-2}$\\
$\Gamma$ &--  & 2.18 (Fix)\\
Norm$^{\mathrm{b}}$ & -- & $1.9^{+2.0}_{-1.9}\,(^{+2.0}_{-1.9})\times10^{-4}$ \\
$\chi^2$/dof & 1241/1180 & 1238/1179 \\ \hline
\end{tabular}
\begin{list}{}{}
\item[$^{\mathrm{a}}$] Normalization of the APEC model.
\item[$^{\mathrm{b}}$] Normalization of the power-law model, in units of ${\rm
    photons\,keV^{-1}\,cm^{-2}\,s^{-1}}$ at 1~keV.
\end{list}
\end{table*}
%%%%%%%%%%%%%%%%%%%%%%%%%%%%%%%%%%%%%%%%%%%%

\subsection{Multi-temperature model}\label{subsec:multi-t}
\begin{figure*}[htb]
   \centering
\rotatebox{0}{\scalebox{0.28}{\includegraphics{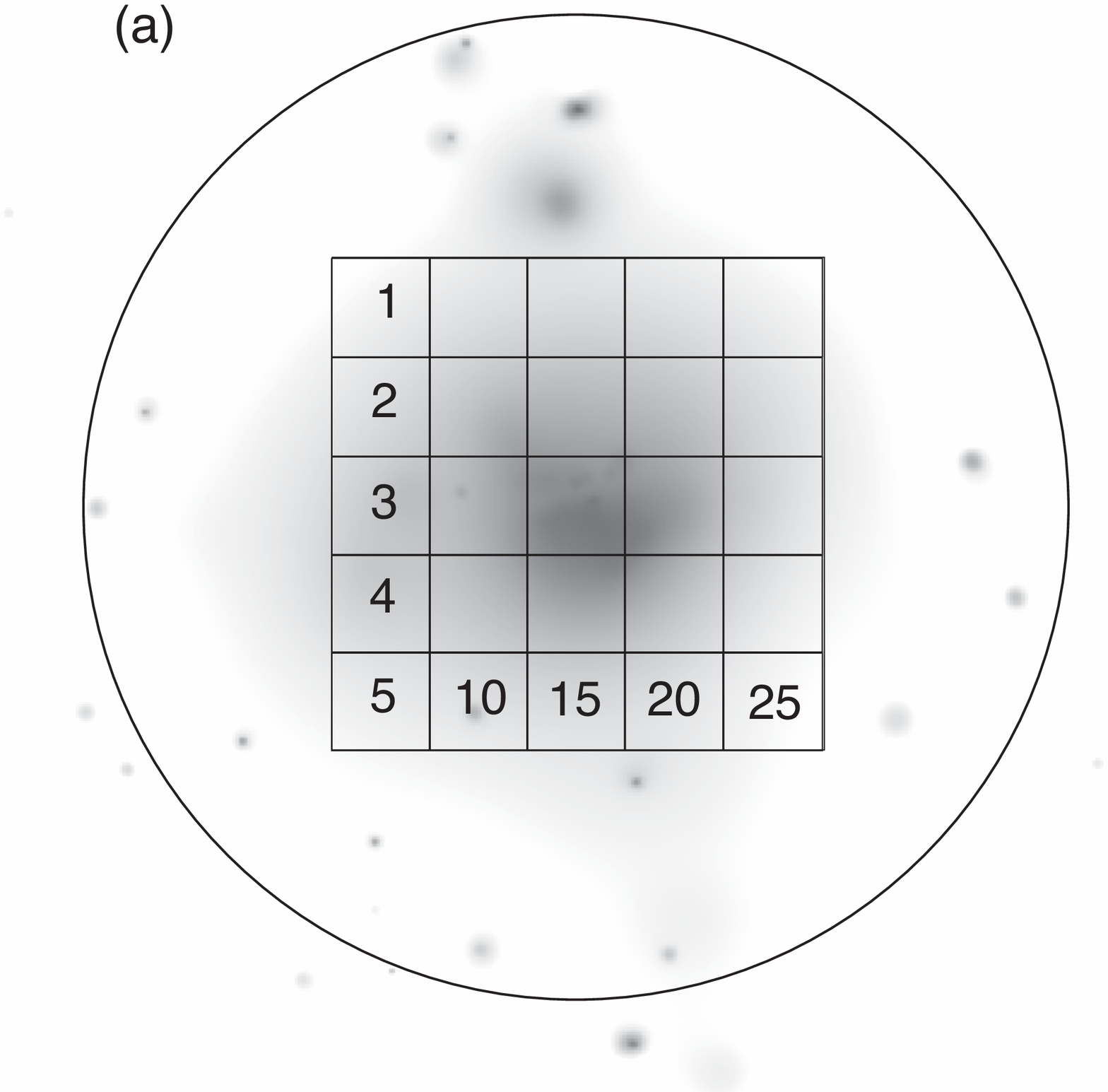}}}
\rotatebox{0}{\scalebox{0.2}{\includegraphics{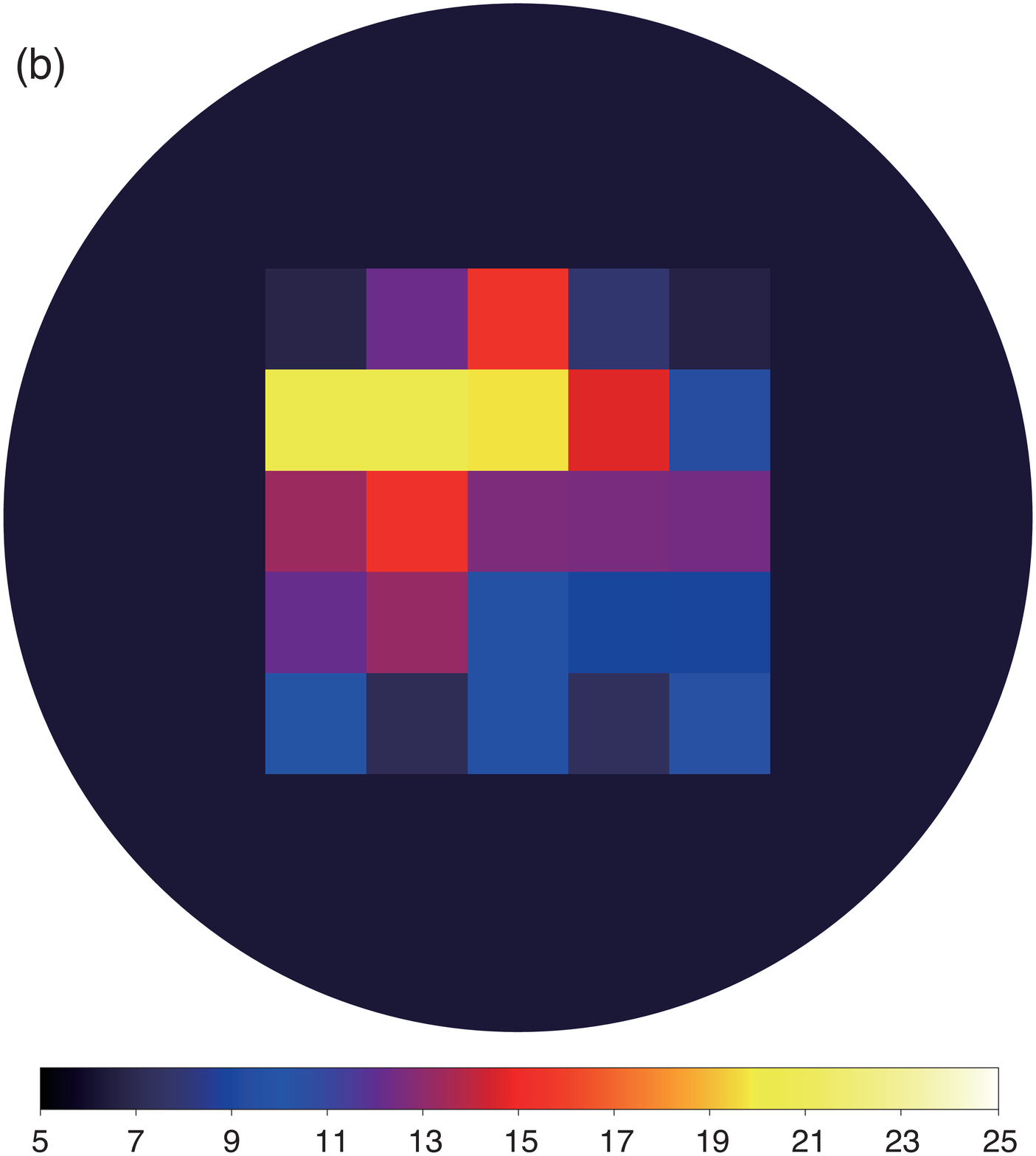}}}
\rotatebox{0}{\scalebox{0.26}{\includegraphics{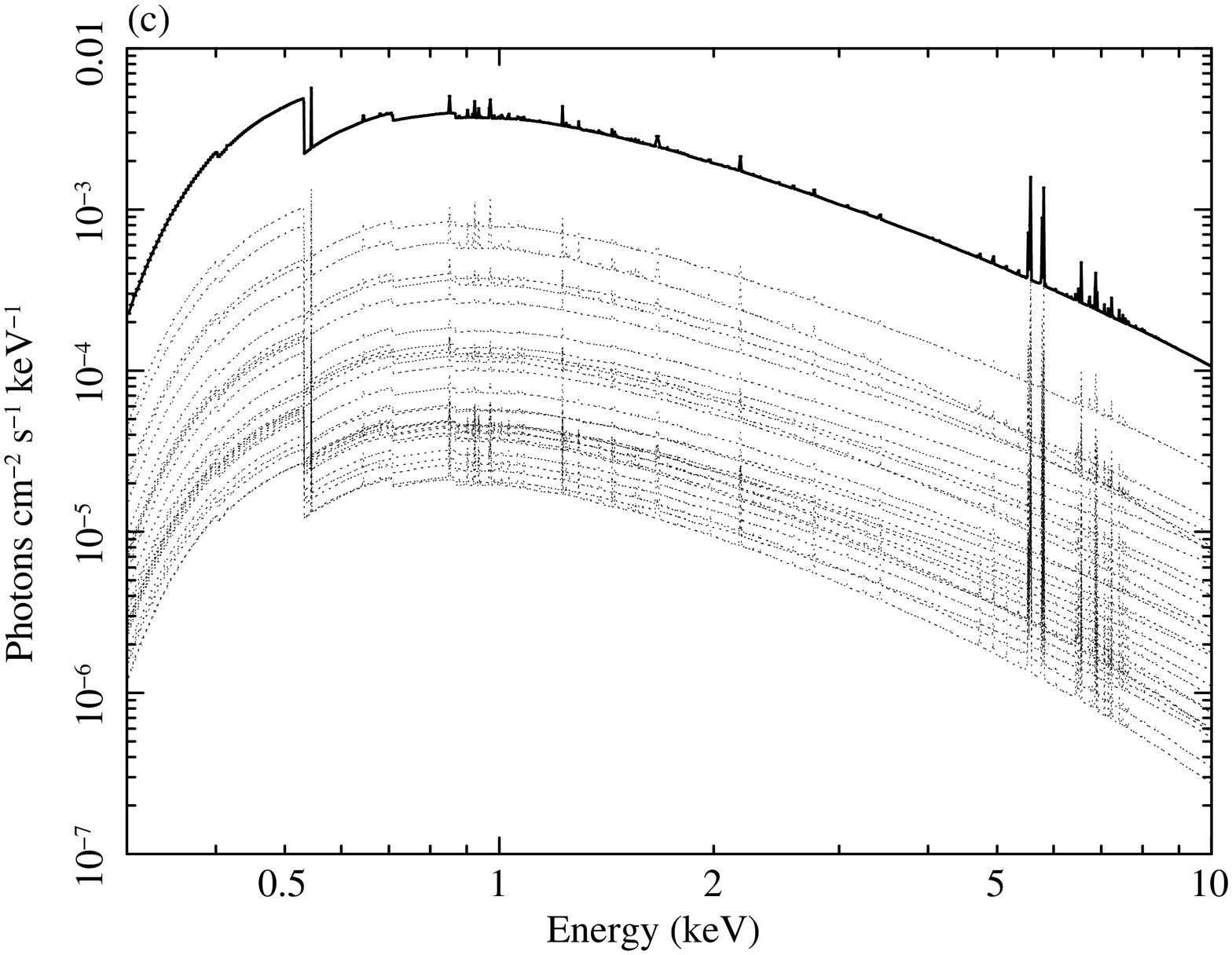}}}
\caption{Multi-temperature model derived from the {\it XMM-Newton} data. (a)
  Definition of spectral regions used to construct the
  multi-temperature model, (b) temperature map obtained from the {\it XMM-Newton}
  spectral analysis, (c) the best-fit multi-temperature model. Panel
  (a) shows the central 25 grids (1 grid = $2\arcmin \times 2\arcmin$)
  and the surrounding region within a circle of $r=10\arcmin$. In
  panel (b), the color scale indicates the gas temperature in keV,
  ranging from 5 (navy) to 25 (white). Panel (c) shows the total model
  (solid line) as well as the APEC models for the 26 spectral regions
  within $r<10\arcmin$ (the dashed lines). }
   \label{fig5}%
 \end{figure*}

Multi-temperature modeling enables more accurate determination of the
hard X-ray property. Because of the complex temperature distribution
in \object{A2163}, the multi-temperature model was constructed from
the spatially-resolved {\it XMM-Newton} EMOS spectra. As shown in
Figure~\ref{fig5}a, the central $10\arcmin\times10\arcmin$ square was
divided into $2\arcmin\times 2\arcmin$ grids. The {\it XMM-Newton}
spectra extracted from the central 25 grids and the surrounding region
inside the circle of $r=10\arcmin$. 26 spectra, represented by a
single-temperature APEC model, were simultaneously fitted under the
condition that their sum reproduces the global cluster spectrum
($r<10\arcmin$) obtained from {\it XMM-Newton} EPN.  The APEC
normalization factor and the temperature were freely selectable within
each region, but the metal abundance was fixed at its mean value
(0.29~solar) for all regions.

Panels b and c of Fig.~\ref{fig5} show the resultant temperature map
and the best-fit multi-temperature spectral model, respectively. The
observed EPN spectra were well represented by the multi-temperature
model. The temperature deduced from this analysis showed that very hot
($kT\sim 18$~keV) gas exists in the North-East region (regions 2, 7,
12 in Fig.~\ref{fig5}a), consistent with
\citet{2011A&A...527A..21B}. Based on this model, the
absorption-corrected energy flux of the very hot gas is estimated as
$F_{\rm NE} = 5.4\times10^{-12}$ and $2.7\times10^{-12}~{\rm
  erg\,s^{-1}cm^{-2}}$ in the 0.5--10, 12--60~keV bands,
respectively. The impact of uncertainty in the hard X-ray flux emitted
by the very hot gas will be examined later.

Next, to investigate whether it properly fits the observed HXD
spectrum, the multi-temperature model was extrapolated to the hard
X-ray band. Since the APEC normalization factors of the HXD and XMM
data are consistent (\S\ref{sec:hxd_analysis}), the relative
normalization was fixed at 1. The reduced $\chi^2$ was 3.3, 8.8, and
0.9 for NXB rescaling factors of 1.00, 0.98, and 1.02,
respectively. Thus the goodness of fit is sensitive to the amplitude
of the NXB model.

\begin{figure}[htb]
   \centering
\rotatebox{0}{\scalebox{0.33}{\includegraphics{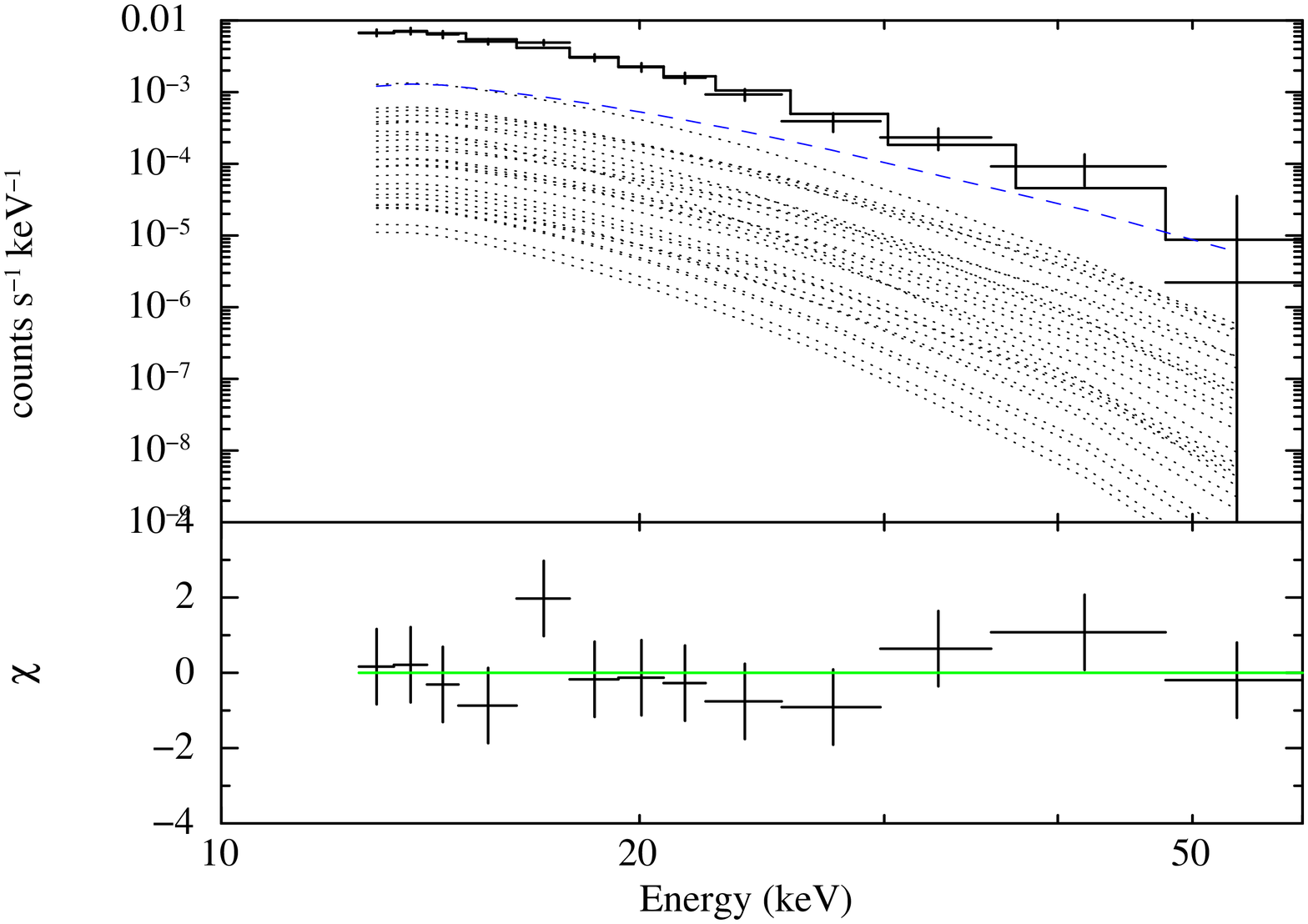}}}
\caption{HXD spectra of \object{A2163} in the 12--60~keV band
  (crosses) fitted by the multi-temperature + power-law model. The
  solid line is the best-fit model. Spectral components of the
  multi-temperature and power-law models are shown as black dotted and
  blue dashed lines, respectively.}
   \label{fig6}%
 \end{figure}

To constrain IC emission, the HXD-PIN data was then fitted by a model
consisting of thermal and non-thermal power-law models
(Fig.~\ref{fig6}). The thermal component was fixed to the best-fit
multi-temperature model derived above, while the power-law
normalization was unrestrained. In the multi-temperature model, the
relative normalization factor between HXD and XMM was again fixed to
1.0. In this model, the chi-squared value is 8.0 for 12 degrees of
freedom. The estimated 12--60~keV power-law flux is $F_{\rm NT} =
5.3\pm0.9\,(\pm3.8)\times10^{-12}~{\rm erg\,s^{-1}cm^{-2}}$ (where the
first and second errors are the $1\sigma$ statistical error and
$1\sigma$ NXB systematic error, respectively). The fitting results and
power-law fluxes after removing the NXB model (rescaled by factors of
1 and 0.98) are summarized in Table~\ref{tab4}.

Another possible source of systematic error is flux uncertainty in the
high-temperature component of the NE region, whose hard X-ray emission
contributes approximately 15\% to the observed HXD spectrum. Fitting
the APEC model to the EMOS spectra accumulated from regions 2, 7, and
12, the 12--60~keV flux and its $1\sigma$ error was estimated to be
$(2.7\pm0.5)\times10^{-12}~{\rm erg\,s^{-1}\,cm^{-2}}$. Thus, the flux
uncertainty of the hot component is less than the systematic error in
NXB. Adding the statistical and systematic errors in quadrature, the
12--60~keV power-law flux is obtained as $(5.3\pm3.9) \times
10^{-12}~{\rm erg\,s^{-1}cm^{-2}}$.  Therefore, IC emission is only
marginally detected ($1.3\sigma$); its 90\% upper limit is $F_{\rm NT}
< 1.2 \times10^{-11}~{\rm erg\,s^{-1}}$ for $\Gamma=2.18$.  Assuming
$\Gamma=1.5$, the non-thermal 12--60~keV flux is $F_{\rm NT}<
1.6\times 10^{-11}~{\rm erg\,s^{-1}}$, corresponding to
$1.7\times10^{-11}~{\rm erg\,s^{-1}}$ in the 20--80~keV.  Thus, the
accuracy of our joint analysis is improved three fold from that of
previous long {\it RXTE} observations \citep[$F_{\rm NT}(20-80~{\rm
  keV})\sim 1.1^{+1.7}_{-0.9}\times10^{-11}~{\rm
  erg\,s^{-1}cm^{-2}}$;][]{2006ApJ...649..673R} and a stronger limit
on non-thermal emission is imposed.

\object{A2163} and \object{RX~J1347.5--1145}
\citep{2008A&A...491..363O} represent the sole examples for which
strong constraints on non-thermal emission have been derived from a
detailed multi-temperature broad-band spectral analysis. As
demonstrated here, the joint analysis allows us to take advantage of
both Suzaku's high spectral capability in the hard X-ray band and
XMM's spatial resolution, and it is worth applying to other merging
clusters to improve the precision of the hard X-ray measurement of the
non-thermal property.

%%%%%%%%%%%%%%%%%%%%%%%%%%%%%%%%%%%%%%%%%%%%
%%%%%%%%%%%%%%%%%%%%%%%%%%%%%%%%%%%%%%%%%%%%

%%%%%%%%%%%%%%%%%%%%%%%%%%%%%%%%%%%%%%%%%%%%
%%%%%%%%%%%%%%%%%%%%%%%%%%%%%%%%%%%%%%%%%%%%
%%%%%%%%%%%%%%%%%%%%%%%%%%%%%%%%%%%%%%%%%%%%
\begin{table*}
  \caption{Result of fitting the multi-temperature + power-law model to the HXD data}\label{tab4}
\centering
\begin{tabular}{lllllll} \hline\hline
 $\Gamma$ & \multicolumn{2}{c}{Norm$^{\mathrm{a}}$} & \multicolumn{2}{c}{$\chi^2/{\rm dof}$} & Flux$^{\mathrm{b}}$ & 90\% upper limit$^{\mathrm{b}}$ \\ \cline{2-3} \cline{4-5}
 & NXB$\times 1.00$ & NXB$\times 0.98$ & NXB$\times 1.00$ & NXB$\times 0.98$ & [${\rm erg\,s^{-1}cm^{-2}}$] & \\ \hline
2.18 & $3.7^{+1.0}_{-1.0}\times10^{-3}$ & $6.4\times10^{-3}$ & 8.0/12 & 11.9/12 & $5.3\pm0.9\,(\pm 3.8)\times10^{-12}$ & ($<1.2\times10^{-11}$) \\
 1.50 & $5.0^{+1.4}_{-1.4}\times10^{-4}$ & $8.7\times 10^{-4}$ & 9.1/12 & 11.8/12 & $6.9\pm1.2\,(\pm 5.1)\times10^{-12}$& ($<1.6\times10^{-11}$) \\ \hline
\end{tabular}
\begin{list}{}{}
\item[$^{\mathrm{a}}$] Normalization of the power-law model, in units of ${\rm photons\,keV^{-1}\,cm^{-2}\,s^{-1}}$ at 1~keV.
\item[$^{\mathrm{b}}$] The 12--60~keV flux of the power-law component and the 90\% upper limit (${\rm erg\,s^{-1}cm^{-2}}$)
\end{list}
\end{table*}
%%%%%%%%%%%%%%%%%%%%%%%%%%%%%%%%%%%%%%%%%%%%

\section{Discussion and conclusion}
Analyzing the long {\it Suzaku} HXD observations, we detected
significant hard X-ray emission from the hottest Abell cluster
\object{A2163} at $z=0.2$. In \S\ref{subsec:origin} we discuss the
origin of this hard X-ray emission and compare the results with
previous observations of other clusters. In \S\ref{subsec:mag}, we
estimate the magnetic field in the cluster by comparing the hard X-ray
flux with radio synchrotron emission.

\subsection{Origin of hard X-ray emission from \object{A2163}}\label{subsec:origin}
Analyzing the high-quality data collected by {\it Suzaku} HXD, we find
that the hard X-ray emission from \object{A2163} is well approximated
by the 14~keV thermal model.  Because \object{A2163} has a complex
temperature distribution, the multi-temperature model (including the
very hot ($kT=18$~keV) gas in the North-East region) imposes more
accurate constraint on the non-thermal flux.  Following a careful
assessment of the uncertainty in the non-X-ray background, we obtained
a 90\% upper limit of $F_{\rm NT} <1.2\times10^{-11}~{\rm
  erg\,s^{-1}cm^{-2}}$ due to inverse Compton emission in the 12--60
keV band. Thus the present result, even though it places tighter
constraints on the non-thermal flux than previous analyses, still
suggests that the emission in this energy band is predominantly of
thermal origin.

It is worth noting that the very hot ($kT\sim18$~keV) gas in the
North-East region contributes non-negligibly to the hard X-ray
emission. In fact, the very hot gas contributes to 15\% of the
observed HXD flux. This finding is reminiscent of the hottest
($kT\sim25$~keV) gas in the distant merging cluster
\object{RX~J1347.5--1145} reported by
\citet{2008A&A...491..363O}. They suggested that such high-temperature
gas is over-pressurized, and will therefore disappear over a
relatively short time scale ($0.5$~Giga years). The presence of very
hot gas in \object{A2163} supports a recent merger of this cluster, as
also suggested from multi-wavelength observations.

\begin{figure}[htb]
   \centering
\rotatebox{0}{\scalebox{0.35}{\includegraphics{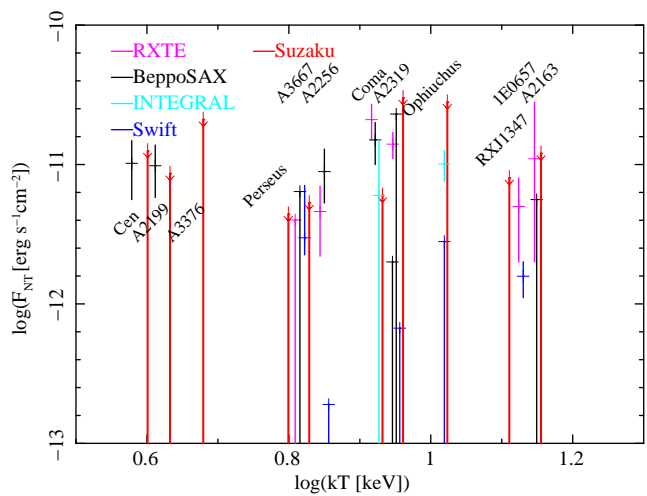}}}
\caption{Non-thermal hard X-ray flux for 12 clusters, measured by {\it
    RXTE} (magenta), {\it Beppo-SAX} (black), {\it INTEGRAL} (light
  blue), {\it Swift} (blue), and {\it Suzaku} (red). {\it RXTE} and
  {\it Beppo-SAX} results are taken from
  \citet[][]{2008SSRv..tmp...16R} and references therein.  The {\it
    INTEGRAL} results are taken from
  \citet{2008ApJ...687..968L,2008A&A...479...27E}.  The {\it Swift}
  results are taken from
  \citet{2009ApJ...690..367A,2010ApJ...725.1688A,2011ApJ...727..119W}. The
  {\it Suzaku} results are presented in
  \citet{2007xnnd.confP..28K,2008PASJ...60.1133F,2008A&A...491..363O,2009PASJ...61S.377K,2009PASJ...61.1293S,2009PASJ...61..339N,2009ApJ...696.1700W,2010PASJ...62....9N,2010PASJ...62..115K}.}
   \label{fig7}%
 \end{figure}

Next, we compare the obtained limit on the IC flux in \object{A2163}
with that of other clusters. \citet{2012RAA....12..973O} compiled the
IC flux measurements from several observatories with hard X-ray
capability, namely, {\it RXTE}, {\it Beppo-SAX}, {\it INTEGRAL}, {\it
  Swift} and {\it Suzaku}.  Fig.~\ref{fig7} shows the fluxes of 12
clusters as a function of gas temperature.  As seen in this figure,
different measurements yielded different fluxes, although their error
bars overlapped for most of the objects.  Since IC emission cannot be
confirmed from these measurements, independent experiments are
required. It should be noted that the IC flux estimation strongly
depends on modeling of both thermal component and the power-law index
of the non-thermal component. Furthermore, the published results are
based on different assumptions. To further explore shock heating and
particle acceleration in clusters, a higher sensitivity in the hard
X-ray band is needed.

%%%%%%%%%%%%%%%%%%%%%%%%%%%%%%%%%%%%%%%%%%%%
%%%%%%%%%%%%%%%%%%%%%%%%%%%%%%%%%%%%%%%%%%%%

\subsection{Cluster magnetic field}\label{subsec:mag}
\cite{1970RvMP...42..237B} derived equations for the radio
  synchrotron emission at the frequency $\nu_{\rm Syn}$ and the IC
  hard X-ray emission at $\nu_{\rm IC}$:
\begin{eqnarray}
\frac{dW_{\rm Syn}}{d\nu_{\rm Syn} dt} & =& \frac{4\pi N_0 e^3 B^{(p+1)/2}}{m_e c^2}
\left(\frac{3e}{4\pi m_e c }\right)^{(p-1)/2}a(p)\nu_{\rm Syn}^{-(p-1)/2}, \label{eq4}\\
\frac{dW_{\rm IC}}{d\nu_{\rm IC} dt} &=& \frac{8\pi^2 r_0^2}{c^2}h^{-(p+3)/2}N_0(k T_{\rm CMB})^{(p+5)/2}F(p)\nu_{\rm IC}^{-(p-1)/2}, \label{eq5}
\end{eqnarray}
where $N_0$ and $p$ are the normalization and the power-law index of
the electron distribution, $N(\gamma)=N_0 \gamma^{-p}$ ($\gamma$ is
the Lorentz factor of the electron), $r_0$ is the classical electron
radius, $h$ is the Planck constant, $T_{\rm CMB}$ is CMB temperature,
and $T_{\rm CMB}=2.73(1+z)$~K.  $a(p)$ and $F(p)$ are given by
Eqs. (4.60) and (2.66) in \cite{1970RvMP...42..237B}, respectively.
Given that the ratio of observed flux densities of the IC hard X-ray
emission $S_{\rm IC}$ to the radio synchrotron emission $S_{\rm Syn}$
is equal to $(dW_{\rm Syn}/d\nu_{\rm Syn} dt)/(dW_{\rm IC}/d\nu_{\rm
  IC} dt)$, the magnetic field in the intracluster space $B$ can be
directly estimated.

Substituting $S_{\rm IC}< 0.25~{\rm \mu Jy}$ at 12~keV (derived from
the joint analysis with $\Gamma=2.18$ (\S\ref{subsec:multi-t})) and
$S_{\rm syn}=155$~mJy at 1.4~GHz \citep{2004A&A...423..111F}, the 90\%
lower limit on the magnetic field in \object{A2163} is obtained as
$B>0.098~{\rm \mu G}$. When $\Gamma=1.5$, we obtain $B>0.006~{\rm \mu
  G}$. The above limits, though weak, are consistent with those of
other clusters, $B\sim 0.1-1~{\rm \mu G}$
\citep[e.g.,][]{2008SSRv..tmp...16R,2010ApJ...725.1688A}.

While our results provide important information on the non-thermal
nature of ICM, their accuracy is limited by the sensitivity of the
hard X-ray instrument. We expect that hard X-ray imaging by {\it
  NuSTAR} \citep{2013ApJ...770..103H} and the future {\it ASTRO-H}
mission \citep{2012SPIE.8443E..1ZT} will enable more accurate
determination of very hot thermal components that predominate in the
hard X-ray band. In addition, these instruments should accurately
locate the merger shock, thereby improving the signal-to-noise ratio
of the non-thermal component.

\begin{acknowledgements}
  The authors thank the {\it Suzaku} team for the operation and
  instrumental calibrations. We also thank R. Smith and H. Yamaguchi
  for providing the plasma code, which covers a broader energy
  range. This work is supported in part by the Grant-in-Aid by MEXT,
  KAKENHI Grant Number 22740124 (NO). 
  GWP acknowledges ANR grant ANR-11-BD56-015.
  THR acknowledges support by the
  Deutsche Forschungsgemeinschaft (DFG) through Emmy Noether grant RE
  1462/2, Heisenberg grant RE 1462/5 and grant RE 1462/6.
\end{acknowledgements}

\bibliographystyle{aa}
\bibliography{a2163_aa}

\begin{thebibliography}{44}
\expandafter\ifx\csname natexlab\endcsname\relax\def\natexlab#1{#1}\fi

\bibitem[{{Ajello} {et~al.}(2009){Ajello}, {Rebusco}, {Cappelluti}, {Reimer},
  {B{\"o}hringer}, {Greiner}, {Gehrels}, {Tueller}, \&
  {Moretti}}]{2009ApJ...690..367A}
{Ajello}, M., {Rebusco}, P., {Cappelluti}, N., {et~al.} 2009, \apj, 690, 367

\bibitem[{{Ajello} {et~al.}(2010){Ajello}, {Rebusco}, {Cappelluti}, {Reimer},
  {B{\"o}hringer}, {La Parola}, \& {Cusumano}}]{2010ApJ...725.1688A}
{Ajello}, M., {Rebusco}, P., {Cappelluti}, N., {et~al.} 2010, \apj, 725, 1688

\bibitem[{{Arnaud} {et~al.}(1992){Arnaud}, {Hughes}, {Forman}, {Jones},
  {Lachieze-Rey}, {Yamashita}, \& {Hatsukade}}]{1992ApJ...390..345A}
{Arnaud}, M., {Hughes}, J.~P., {Forman}, W., {et~al.} 1992, \apj, 390, 345

\bibitem[{{Arnaud} {et~al.}(2002){Arnaud}, {Majerowicz}, {Lumb}, {Neumann},
  {Aghanim}, {Blanchard}, {Boer}, {Burke}, {Collins}, {Giard}, {Nevalainen},
  {Nichol}, {Romer}, \& {Sadat}}]{arn02}
{Arnaud}, M., {Majerowicz}, S., {Lumb}, D., {et~al.} 2002, \aap, 390, 27

\bibitem[{{Blumenthal} \& {Gould}(1970)}]{1970RvMP...42..237B}
{Blumenthal}, G.~R. \& {Gould}, R.~J. 1970, Reviews of Modern Physics, 42, 237

\bibitem[{{Boldt}(1987)}]{1987IAUS..124..611B}
{Boldt}, E. 1987, in IAU Symposium, Vol. 124, Observational Cosmology, ed.
  A.~{Hewitt}, G.~{Burbidge}, \& L.~Z. {Fang}, 611--615

\bibitem[{{Bourdin} {et~al.}(2011){Bourdin}, {Arnaud}, {Mazzotta}, {Pratt},
  {Sauvageot}, {Martino}, {Maurogordato}, {Cappi}, {Ferrari}, \&
  {Benoist}}]{2011A&A...527A..21B}
{Bourdin}, H., {Arnaud}, M., {Mazzotta}, P., {et~al.} 2011, \aap, 527, A21

\bibitem[{{Brunetti} {et~al.}(2009){Brunetti}, {Cassano}, {Dolag}, \&
  {Setti}}]{2009A&A...507..661B}
{Brunetti}, G., {Cassano}, R., {Dolag}, K., \& {Setti}, G. 2009, \aap, 507, 661

\bibitem[{{Cassano} {et~al.}(2013){Cassano}, {Ettori}, {Brunetti},
  {Giacintucci}, {Pratt}, {Venturi}, {Kale}, {Dolag}, \&
  {Markevitch}}]{2013arXiv1306.4379C}
{Cassano}, R., {Ettori}, S., {Brunetti}, G., {et~al.} 2013, ArXiv e-prints

\bibitem[{{Eckert} {et~al.}(2008){Eckert}, {Produit}, {Paltani}, {Neronov}, \&
  {Courvoisier}}]{2008A&A...479...27E}
{Eckert}, D., {Produit}, N., {Paltani}, S., {Neronov}, A., \& {Courvoisier},
  T.~J.-L. 2008, \aap, 479, 27

\bibitem[{{Feretti} {et~al.}(2001){Feretti}, {Fusco-Femiano}, {Giovannini}, \&
  {Govoni}}]{2001A&A...373..106F}
{Feretti}, L., {Fusco-Femiano}, R., {Giovannini}, G., \& {Govoni}, F. 2001,
  \aap, 373, 106

\bibitem[{{Feretti} {et~al.}(2012){Feretti}, {Giovannini}, {Govoni}, \&
  {Murgia}}]{2012A&ARv..20...54F}
{Feretti}, L., {Giovannini}, G., {Govoni}, F., \& {Murgia}, M. 2012, \aapr, 20,
  54

\bibitem[{{Feretti} {et~al.}(2004){Feretti}, {Orr{\`u}}, {Brunetti},
  {Giovannini}, {Kassim}, \& {Setti}}]{2004A&A...423..111F}
{Feretti}, L., {Orr{\`u}}, E., {Brunetti}, G., {et~al.} 2004, \aap, 423, 111

\bibitem[{{Fujita} {et~al.}(2008){Fujita}, {Hayashida}, {Nagai}, {Inoue},
  {Matsumoto}, {Okabe}, {Reiprich}, {Sarazin}, \&
  {Takizawa}}]{2008PASJ...60.1133F}
{Fujita}, Y., {Hayashida}, K., {Nagai}, M., {et~al.} 2008, \pasj, 60, 1133

\bibitem[{{Fukazawa} {et~al.}(2009){Fukazawa}, {Mizuno}, {Watanabe}, {Kokubun},
  {Takahashi}, {Kawano}, {Nishino}, {Sasada}, {Shirai}, {Takahashi}, {Umeki},
  {Yamasaki}, {Yasuda}, {Bamba}, {Ohno}, {Takahashi}, {Ushio}, {Enoto},
  {Kitaguchi}, {Makishima}, {Nakazawa}, {Uehara}, {Yamada}, {Yuasa}, {Isobe},
  {Kawaharada}, {Tanaka}, {Tashiro}, {Terada}, \&
  {Yamaoka}}]{2009PASJ...61S..17F}
{Fukazawa}, Y., {Mizuno}, T., {Watanabe}, S., {et~al.} 2009, \pasj, 61, 17

\bibitem[{{Fusco-Femiano} {et~al.}(2011){Fusco-Femiano}, {Orlandini},
  {Bonamente}, \& {Lapi}}]{2011ApJ...732...85F}
{Fusco-Femiano}, R., {Orlandini}, M., {Bonamente}, M., \& {Lapi}, A. 2011,
  \apj, 732, 85

\bibitem[{{Fusco-Femiano} {et~al.}(2004){Fusco-Femiano}, {Orlandini},
  {Brunetti}, {Feretti}, {Giovannini}, {Grandi}, \&
  {Setti}}]{2004ApJ...602L..73F}
{Fusco-Femiano}, R., {Orlandini}, M., {Brunetti}, G., {et~al.} 2004, \apjl,
  602, L73

\bibitem[{{Harrison} {et~al.}(2013){Harrison}, {Craig}, {Christensen},
  {Hailey}, {Zhang}, {Boggs}, {Stern}, {Cook}, {Forster}, {Giommi},
  {Grefenstette}, {Kim}, {Kitaguchi}, {Koglin}, {Madsen}, {Mao}, {Miyasaka},
  {Mori}, {Perri}, {Pivovaroff}, {Puccetti}, {Rana}, {Westergaard}, {Willis},
  {Zoglauer}, {An}, {Bachetti}, {Barri{\`e}re}, {Bellm}, {Bhalerao},
  {Brejnholt}, {Fuerst}, {Liebe}, {Markwardt}, {Nynka}, {Vogel}, {Walton},
  {Wik}, {Alexander}, {Cominsky}, {Hornschemeier}, {Hornstrup}, {Kaspi},
  {Madejski}, {Matt}, {Molendi}, {Smith}, {Tomsick}, {Ajello}, {Ballantyne},
  {Balokovi{\'c}}, {Barret}, {Bauer}, {Blandford}, {Brandt}, {Brenneman},
  {Chiang}, {Chakrabarty}, {Chenevez}, {Comastri}, {Dufour}, {Elvis}, {Fabian},
  {Farrah}, {Fryer}, {Gotthelf}, {Grindlay}, {Helfand}, {Krivonos}, {Meier},
  {Miller}, {Natalucci}, {Ogle}, {Ofek}, {Ptak}, {Reynolds}, {Rigby},
  {Tagliaferri}, {Thorsett}, {Treister}, \& {Urry}}]{2013ApJ...770..103H}
{Harrison}, F.~A., {Craig}, W.~W., {Christensen}, F.~E., {et~al.} 2013, \apj,
  770, 103

\bibitem[{{Huber} {et~al.}(2013){Huber}, {Tchernin}, {Eckert}, {Farnier},
  {Manalaysay}, {Straumann}, \& {Walter}}]{2013A&A...560A..64H}
{Huber}, B., {Tchernin}, C., {Eckert}, D., {et~al.} 2013, \aap, 560, A64

\bibitem[{{Kawaharada} {et~al.}(2010){Kawaharada}, {Makishima}, {Kitaguchi},
  {Okuyama}, {Nakazawa}, \& {Fukazawa}}]{2010PASJ...62..115K}
{Kawaharada}, M., {Makishima}, K., {Kitaguchi}, T., {et~al.} 2010, \pasj, 62,
  115

\bibitem[{{Kawano} {et~al.}(2009){Kawano}, {Fukazawa}, {Nishino}, {Nakazawa},
  {Kitaguchi}, {Makishima}, {Takahashi}, {Kokubun}, {Ota}, {Ohashi}, {Isobe},
  {Henry}, \& {Hornschemeier}}]{2009PASJ...61S.377K}
{Kawano}, N., {Fukazawa}, Y., {Nishino}, S., {et~al.} 2009, \pasj, 61, 377

\bibitem[{{Kitaguchi} {et~al.}(2007){Kitaguchi}, {Nakazawa}, {Makishima},
  {Kawaharada}, {Ota}, {Kokubun}, {Yamasaki}, {Kawano}, {Fukazawa}, {Sato},
  {Ohashi}, {Murase}, {Urata}, {Tashiro}, {Furusawa}, \& {Suzaku
  Team}}]{2007xnnd.confP..28K}
{Kitaguchi}, T., {Nakazawa}, N., {Makishima}, K., {et~al.} 2007, in XMM-Newton:
  The Next Decade, 28P

\bibitem[{{Lutovinov} {et~al.}(2008){Lutovinov}, {Vikhlinin}, {Churazov},
  {Revnivtsev}, \& {Sunyaev}}]{2008ApJ...687..968L}
{Lutovinov}, A.~A., {Vikhlinin}, A., {Churazov}, E.~M., {Revnivtsev}, M.~G., \&
  {Sunyaev}, R.~A. 2008, \apj, 687, 968

\bibitem[{{Markevitch} \& {Vikhlinin}(2001)}]{2001ApJ...563...95M}
{Markevitch}, M. \& {Vikhlinin}, A. 2001, \apj, 563, 95

\bibitem[{{Mitsuda} {et~al.}(2007){Mitsuda}, {Bautz}, {Inoue}, {Kelley},
  {Koyama}, {Kunieda}, {Makishima}, {Ogawara}, {Petre}, {Takahashi}, {Tsunemi},
  {White}, {Anabuki}, {Angelini}, {Arnaud}, {Awaki}, {Bamba}, {Boyce}, {Brown},
  {Chan}, {Cottam}, {Dotani}, {Doty}, {Ebisawa}, {Ezoe}, {Fabian}, {Figueroa},
  {Fujimoto}, {Fukazawa}, {Furusho}, {Furuzawa}, {Gendreau}, {Griffiths},
  {Haba}, {Hamaguchi}, {Harrus}, {Hasinger}, {Hatsukade}, {Hayashida}, {Henry},
  {Hiraga}, {Holt}, {Hornschemeier}, {Hughes}, {Hwang}, {Ishida}, {Ishisaki},
  {Isobe}, {Itoh}, {Iyomoto}, {Kahn}, {Kamae}, {Katagiri}, {Kataoka},
  {Katayama}, {Kawai}, {Kilbourne}, {Kinugasa}, {Kissel}, {Kitamoto}, {Kohama},
  {Kohmura}, {Kokubun}, {Kotani}, {Kotoku}, {Kubota}, {Madejski}, {Maeda},
  {Makino}, {Markowitz}, {Matsumoto}, {Matsumoto}, {Matsuoka}, {Matsushita},
  {McCammon}, {Mihara}, {Misaki}, {Miyata}, {Mizuno}, {Mori}, {Mori}, {Morii},
  {Moseley}, {Mukai}, {Murakami}, {Murakami}, {Mushotzky}, {Nagase}, {Namiki},
  {Negoro}, {Nakazawa}, {Nousek}, {Okajima}, {Ogasaka}, {Ohashi}, {Oshima},
  {Ota}, {Ozaki}, {Ozawa}, {Parmar}, {Pence}, {Porter}, {Reeves}, {Ricker},
  {Sakurai}, {Sanders}, {Senda}, {Serlemitsos}, {Shibata}, {Soong}, {Smith},
  {Suzuki}, {Szymkowiak}, {Takahashi}, {Tamagawa}, {Tamura}, {Tamura},
  {Tanaka}, {Tashiro}, {Tawara}, {Terada}, {Terashima}, {Tomida}, {Torii},
  {Tsuboi}, {Tsujimoto}, {Tsuru}, {Turner}, {Ueda}, {Ueno}, {Ueno}, {Uno},
  {Urata}, {Watanabe}, {Yamamoto}, {Yamaoka}, {Yamasaki}, {Yamashita},
  {Yamauchi}, {Yamauchi}, {Yaqoob}, {Yonetoku}, \&
  {Yoshida}}]{2007PASJ...59S...1M}
{Mitsuda}, K., {Bautz}, M., {Inoue}, H., {et~al.} 2007, \pasj, 59, 1

\bibitem[{{Nakazawa} {et~al.}(2009){Nakazawa}, {Sarazin}, {Kawaharada},
  {Kitaguchi}, {Okuyama}, {Makishima}, {Kawano}, {Fukazawa}, {Inoue},
  {Takizawa}, {Wik}, {Finoguenov}, \& {Clarke}}]{2009PASJ...61..339N}
{Nakazawa}, K., {Sarazin}, C.~L., {Kawaharada}, M., {et~al.} 2009, \pasj, 61,
  339

\bibitem[{{Nishino} {et~al.}(2010){Nishino}, {Fukazawa}, {Hayashi}, {Nakazawa},
  \& {Tanaka}}]{2010PASJ...62....9N}
{Nishino}, S., {Fukazawa}, Y., {Hayashi}, K., {Nakazawa}, K., \& {Tanaka}, T.
  2010, \pasj, 62, 9

\bibitem[{{Okabe} {et~al.}(2011){Okabe}, {Bourdin}, {Mazzotta}, \&
  {Maurogordato}}]{2011ApJ...741..116O}
{Okabe}, N., {Bourdin}, H., {Mazzotta}, P., \& {Maurogordato}, S. 2011, \apj,
  741, 116

\bibitem[{{Ota}(2012)}]{2012RAA....12..973O}
{Ota}, N. 2012, Research in Astronomy and Astrophysics, 12, 973

\bibitem[{{Ota} {et~al.}(2008){Ota}, {Murase}, {Kitayama}, {Komatsu},
  {Hattori}, {Matsuo}, {Oshima}, {Suto}, \& {Yoshikawa}}]{2008A&A...491..363O}
{Ota}, N., {Murase}, K., {Kitayama}, T., {et~al.} 2008, \aap, 491, 363

\bibitem[{Pratt {et~al.}(2007)Pratt, B{\"o}hringer, Croston, Arnaud, Borgani,
  Finoguenov, \& Temple}]{pra07}
Pratt, G.~W., B{\"o}hringer, H., Croston, J.~H., {et~al.} 2007, \aap, 461, 71

\bibitem[{{Read} \& {Ponman}(2003)}]{rea03}
{Read}, A.~M. \& {Ponman}, T.~J. 2003, \aap, 409, 395

\bibitem[{{Rephaeli} \& {Gruber}(2002)}]{2002ApJ...579..587R}
{Rephaeli}, Y. \& {Gruber}, D. 2002, \apj, 579, 587

\bibitem[{{Rephaeli} {et~al.}(2006){Rephaeli}, {Gruber}, \&
  {Arieli}}]{2006ApJ...649..673R}
{Rephaeli}, Y., {Gruber}, D., \& {Arieli}, Y. 2006, \apj, 649, 673

\bibitem[{{Rephaeli} {et~al.}(2008){Rephaeli}, {Nevalainen}, {Ohashi}, \&
  {Bykov}}]{2008SSRv..tmp...16R}
{Rephaeli}, Y., {Nevalainen}, J., {Ohashi}, T., \& {Bykov}, A.~M. 2008, Space
  Science Reviews, 16

\bibitem[{{Smith} {et~al.}(2001){Smith}, {Brickhouse}, {Liedahl}, \&
  {Raymond}}]{2001ApJ...556L..91S}
{Smith}, R.~K., {Brickhouse}, N.~S., {Liedahl}, D.~A., \& {Raymond}, J.~C.
  2001, \apjl, 556, L91

\bibitem[{{Soucail}(2012)}]{2012A&A...540A..61S}
{Soucail}, G. 2012, \aap, 540, A61

\bibitem[{{Sugawara} {et~al.}(2009){Sugawara}, {Takizawa}, \&
  {Nakazawa}}]{2009PASJ...61.1293S}
{Sugawara}, C., {Takizawa}, M., \& {Nakazawa}, K. 2009, \pasj, 61, 1293

\bibitem[{{Takahashi} {et~al.}(2007){Takahashi}, {Abe}, {Endo}, {Endo}, {Ezoe},
  {Fukazawa}, {Hamaya}, {Hirakuri}, {Hong}, {Horii}, {Inoue}, {Isobe}, {Itoh},
  {Iyomoto}, {Kamae}, {Kasama}, {Kataoka}, {Kato}, {Kawaharada}, {Kawano},
  {Kawashima}, {Kawasoe}, {Kishishita}, {Kitaguchi}, {Kobayashi}, {Kokubun},
  {Kotoku}, {Kouda}, {Kubota}, {Kuroda}, {Madejski}, {Makishima}, {Masukawa},
  {Matsumoto}, {Mitani}, {Miyawaki}, {Mizuno}, {Mori}, {Mori}, {Murashima},
  {Murakami}, {Nakazawa}, {Niko}, {Nomachi}, {Okada}, {Ohno}, {Oonuki}, {Ota},
  {Ozawa}, {Sato}, {Shinoda}, {Sugiho}, {Suzuki}, {Taguchi}, {Takahashi},
  {Takahashi}, {Takeda}, {Tamura}, {Tamura}, {Tanaka}, {Tanihata}, {Tashiro},
  {Terada}, {Tominaga}, {Uchiyama}, {Watanabe}, {Yamaoka}, {Yanagida}, \&
  {Yonetoku}}]{2007PASJ...59S..35T}
{Takahashi}, T., {Abe}, K., {Endo}, M., {et~al.} 2007, \pasj, 59, 35

\bibitem[{{Takahashi} {et~al.}(2012){Takahashi}, {Mitsuda}, {Kelley}, {Aarts},
  {Aharonian}, {Akamatsu}, {Akimoto}, {Allen}, {Anabuki}, {Angelini}, {Arnaud},
  {Asai}, {Audard}, {Awaki}, {Azzarello}, {Baluta}, {Bamba}, {Bando}, {Bautz},
  {Blandford}, {Boyce}, {Brown}, {Cackett}, {Chernyakova}, {Coppi},
  {Costantini}, {de Plaa}, {den Herder}, {DiPirro}, {Done}, {Dotani}, {Doty},
  {Ebisawa}, {Eckart}, {Enoto}, {Ezoe}, {Fabian}, {Ferrigno}, {Foster},
  {Fujimoto}, {Fukazawa}, {Funk}, {Furuzawa}, {Galeazzi}, {Gallo}, {Gandhi},
  {Gendreau}, {Gilmore}, {Haas}, {Haba}, {Hamaguchi}, {Hatsukade}, {Hayashi},
  {Hayashida}, {Hiraga}, {Hirose}, {Hornschemeier}, {Hoshino}, {Hughes},
  {Hwang}, {Iizuka}, {Inoue}, {Ishibashi}, {Ishida}, {Ishimura}, {Ishisaki},
  {Ito}, {Iwata}, {Iyomoto}, {Kaastra}, {Kallman}, {Kamae}, {Kataoka},
  {Katsuda}, {Kawahara}, {Kawaharada}, {Kawai}, {Kawasaki}, {Khangaluyan},
  {Kilbourne}, {Kimura}, {Kinugasa}, {Kitamoto}, {Kitayama}, {Kohmura},
  {Kokubun}, {Kosaka}, {Koujelev}, {Koyama}, {Krimm}, {Kubota}, {Kunieda},
  {LaMassa}, {Laurent}, {Lebrun}, {Leutenegger}, {Limousin}, {Loewenstein},
  {Long}, {Lumb}, {Madejski}, {Maeda}, {Makishima}, {Marchand}, {Markevitch},
  {Matsumoto}, {Matsushita}, {McCammon}, {McNamara}, {Miller}, {Miller},
  {Mineshige}, {Minesugi}, {Mitsuishi}, {Miyazawa}, {Mizuno}, {Mori}, {Mori},
  {Mukai}, {Murakami}, {Murakami}, {Mushotzky}, {Nagano}, {Nagino}, {Nakagawa},
  {Nakajima}, {Nakamori}, {Nakazawa}, {Namba}, {Natsukari}, {Nishioka},
  {Nobukawa}, {Nomachi}, {O'Dell}, {Odaka}, {Ogawa}, {Ogawa}, {Ogi}, {Ohashi},
  {Ohno}, {Ohta}, {Okajima}, {Okamoto}, {Okazaki}, {Ota}, {Ozaki}, {Paerels},
  {Paltani}, {Parmar}, {Petre}, {Pohl}, {Porter}, {Ramsey}, {Reis}, {Reynolds},
  {Russell}, {Safi-Harb}, {Sakai}, {Sameshima}, {Sanders}, {Sato}, {Sato},
  {Sato}, {Sato}, {Sawada}, {Serlemitsos}, {Seta}, {Shibano}, {Shida},
  {Shimada}, {Shinozaki}, {Shirron}, {Simionescu}, {Simmons}, {Smith},
  {Sneiderman}, {Soong}, {Stawarz}, {Sugawara}, {Sugita}, {Sugita},
  {Szymkowiak}, {Tajima}, {Takahashi}, {Takeda}, {Takei}, {Tamagawa}, {Tamura},
  {Tamura}, {Tanaka}, {Tanaka}, {Tashiro}, {Tawara}, {Terada}, {Terashima},
  {Tombesi}, {Tomida}, {Tsuboi}, {Tsujimoto}, {Tsunemi}, {Tsuru}, {Uchida},
  {Uchiyama}, {Uchiyama}, {Ueda}, {Ueno}, {Uno}, {Urry}, {Ursino}, {de Vries},
  {Wada}, {Watanabe}, {Werner}, {White}, {Yamada}, {Yamada}, {Yamaguchi},
  {Yamasaki}, {Yamauchi}, {Yamauchi}, {Yatsu}, {Yonetoku}, {Yoshida}, \&
  {Yuasa}}]{2012SPIE.8443E..1ZT}
{Takahashi}, T., {Mitsuda}, K., {Kelley}, R., {et~al.} 2012, in Society of
  Photo-Optical Instrumentation Engineers (SPIE) Conference Series, Vol. 8443,
  Society of Photo-Optical Instrumentation Engineers (SPIE) Conference Series

\bibitem[{{The Fermi-LAT Collaboration} {et~al.}(2013){The Fermi-LAT
  Collaboration}, {:}, {Ackermann}, {Ajello}, {Albert}, {Allafort}, {Atwood},
  {Baldini}, {Ballet}, {Barbiellini}, {Bastieri}, {Bechtol}, {Bellazzini},
  {Bloom}, {Bonamente}, {Bottacini}, {Brandt}, {Bregeon}, {Brigida}, {Bruel},
  {Buehler}, {Buson}, {Caliandro}, {Cameron}, {Caraveo}, {Cavazzuti}, {Chaves},
  {Chiang}, {Chiaro}, {Ciprini}, {Claus}, {Cohen-Tanugi}, {Conrad},
  {D'Ammando}, {de Angelis}, {de Palma}, {Dermer}, {Digel}, {Drell},
  {Drlica-Wagner}, {Favuzzi}, {Franckowiak}, {Funk}, {Fusco}, {Gargano},
  {Gasparrini}, {Germani}, {Giglietto}, {Giordano}, {Giroletti}, {Godfrey},
  {Gomez-Vargas}, {Grenier}, {Guiriec}, {Gustafsson}, {Hadasch}, {Hayashida},
  {Hewitt}, {Hughes}, {Jeltema}, {J{\'o}hannesson}, {Johnson}, {Kamae},
  {Kataoka}, {Kn{\"o}dlseder}, {Kuss}, {Lande}, {Larsson}, {Latronico}, {Llena
  Garde}, {Longo}, {Loparco}, {Lovellette}, {Lubrano}, {Mayer}, {Mazziotta},
  {McEnery}, {Michelson}, {Mitthumsiri}, {Mizuno}, {Monzani}, {Morselli},
  {Moskalenko}, {Murgia}, {Nemmen}, {Nuss}, {Ohsugi}, {Orienti}, {Orlando},
  {Ormes}, {Perkins}, {Pesce-Rollins}, {Piron}, {Pivato}, {Rain{\`o}}, {Rando},
  {Razzano}, {Razzaque}, {Reimer}, {Reimer}, {Ruan}, {S{\'a}nchez-Conde},
  {Schulz}, {Sgr{\`o}}, {Siskind}, {Spandre}, {Spinelli}, {Storm}, {Strong},
  {Suson}, {Takahashi}, {Thayer}, {Thayer}, {Thompson}, {Tibaldo}, {Tinivella},
  {Torres}, {Troja}, {Uchiyama}, {Usher}, {Vandenbroucke}, {Vianello},
  {Vitale}, {Winer}, {Wood}, {Zimmer}, {Pfrommer}, \&
  {Pinzke}}]{2013arXiv1308.5654T}
{The Fermi-LAT Collaboration}, {:}, {Ackermann}, M., {et~al.} 2013, ArXiv
  e-prints

\bibitem[{{Tsujimoto} {et~al.}(2011){Tsujimoto}, {Guainazzi}, {Plucinsky},
  {Beardmore}, {Ishida}, {Natalucci}, {Posson-Brown}, {Read}, {Saxton}, \&
  {Shaposhnikov}}]{2011A&A...525A..25T}
{Tsujimoto}, M., {Guainazzi}, M., {Plucinsky}, P.~P., {et~al.} 2011, \aap, 525,
  A25

\bibitem[{{Wik} {et~al.}(2011){Wik}, {Sarazin}, {Finoguenov}, {Baumgartner},
  {Mushotzky}, {Okajima}, {Tueller}, \& {Clarke}}]{2011ApJ...727..119W}
{Wik}, D.~R., {Sarazin}, C.~L., {Finoguenov}, A., {et~al.} 2011, \apj, 727, 119

\bibitem[{{Wik} {et~al.}(2009){Wik}, {Sarazin}, {Finoguenov}, {Matsushita},
  {Nakazawa}, \& {Clarke}}]{2009ApJ...696.1700W}
{Wik}, D.~R., {Sarazin}, C.~L., {Finoguenov}, A., {et~al.} 2009, \apj, 696,
  1700

\end{thebibliography}

\end{document}